\documentstyle[12pt]{article}
\begin{document}

\def\ds{\displaystyle}
\def\beq{\begin{equation}}
\def\eeq{\end{equation}}
\def\bea{\begin{eqnarray}}
\def\eea{\end{eqnarray}}
\def\beeq{\begin{eqnarray}}
\def\eeeq{\end{eqnarray}}
\def\ve{\vert}
\def\vel{\left|}
\def\ver{\right|}
\def\nnb{\nonumber}
\def\ga{\left(}
\def\dr{\right)}
\def\aga{\left\{}
\def\adr{\right\}}
\def\lla{\left<}
\def\rra{\right>}
\def\rar{\rightarrow}
\def\nnb{\nonumber}
\def\la{\langle}
\def\ra{\rangle}
\def\ba{\begin{array}}
\def\ea{\end{array}}
\def\tr{\mbox{Tr}}
\def\ssp{{\Sigma^{*+}}}
\def\sso{{\Sigma^{*0}}}
\def\ssm{{\Sigma^{*-}}}
\def\xis0{{\Xi^{*0}}}
\def\xism{{\Xi^{*-}}}
\def\qs{\la \bar s s \ra}
\def\qu{\la \bar u u \ra}
\def\qd{\la \bar d d \ra}
\def\qq{\la \bar q q \ra}
\def\gGgG{\la g^2 G^2 \ra}
\def\q{\gamma_5 \not\!q}
\def\x{\gamma_5 \not\!x}
\def\g5{\gamma_5}
\def\sb{S_Q^{cf}}
\def\sd{S_d^{be}}
\def\su{S_u^{ad}}
\def\ss{S_s^{??}}
\def\ll{\Lambda}
\def\lb{\Lambda_b}
\def\sbp{{S}_Q^{'cf}}
\def\sdp{{S}_d^{'be}}
\def\sup{{S}_u^{'ad}}
\def\ssp{{S}_s^{'??}}
\def\sig{\sigma_{\mu \nu} \gamma_5 p^\mu q^\nu}
\def\fo{f_0(\frac{s_0}{M^2})}
\def\ffi{f_1(\frac{s_0}{M^2})}
\def\fii{f_2(\frac{s_0}{M^2})}
\def\O{{\cal O}}
\def\sl{{\Sigma^0 \Lambda}}
\def\es{\!\!\! &=& \!\!\!}
\def\ar{&+& \!\!\!}
\def\ek{&-& \!\!\!}
\def\cp{&\times& \!\!\!}
\def\se{\!\!\! &\simeq& \!\!\!}
\def\hml{\hat{m}_{\ell}}
\def\rr{\hat{r}_{\Lambda}}
\def\ss{\hat{s}}


\renewcommand{\textfraction}{0.2}    
\renewcommand{\topfraction}{0.8}

\renewcommand{\bottomfraction}{0.4}
\renewcommand{\floatpagefraction}{0.8}
\newcommand\mysection{\setcounter{equation}{0}\section}

\def\baeq{\begin{appeq}}     \def\eaeq{\end{appeq}}
\def\baeeq{\begin{appeeq}}   \def\eaeeq{\end{appeeq}}
\newenvironment{appeq}{\beq}{\eeq}
\newenvironment{appeeq}{\beeq}{\eeeq}
\def\bAPP#1#2{
 \markright{APPENDIX #1}
 \addcontentsline{toc}{section}{Appendix #1: #2}
 \medskip
 \medskip
 \begin{center}      {\bf\LARGE Appendix #1 :}{\quad\Large\bf #2}
\end{center}
 \renewcommand{\thesection}{#1.\arabic{section}}
\setcounter{equation}{0}
        \renewcommand{\thehran}{#1.\arabic{hran}}
\renewenvironment{appeq}
  {  \renewcommand{\theequation}{#1.\arabic{equation}}
     \beq
  }{\eeq}
\renewenvironment{appeeq}
  {  \renewcommand{\theequation}{#1.\arabic{equation}}
     \beeq
  }{\eeeq}
\nopagebreak \noindent}

\def\eAPP{\renewcommand{\thehran}{\thesection.\arabic{hran}}}

\renewcommand{\theequation}{\arabic{equation}}
\newcounter{hran}
\renewcommand{\thehran}{\thesection.\arabic{hran}}

\def\bmini{\setcounter{hran}{\value{equation}}
\refstepcounter{hran}\setcounter{equation}{0}
\renewcommand{\theequation}{\thehran\alph{equation}}\begin{eqnarray}}
\def\bminiG#1{\setcounter{hran}{\value{equation}}
\refstepcounter{hran}\setcounter{equation}{-1}
\renewcommand{\theequation}{\thehran\alph{equation}}
\refstepcounter{equation}\label{#1}\begin{eqnarray}}


\newskip\humongous \humongous=0pt plus 1000pt minus 1000pt
\def\caja{\mathsurround=0pt}


\title{
 {\small \begin{flushright}
IPM/P-2007/048\\
\today
\end{flushright}}
       {\Large
                 {\bf
 Double Lepton Polarization Forward--Backward Asymmetries
in $B \rar K^\ast \ell^+ \ell^-$ Decay in the SM4.
                 }
         }
      }

\author{\vspace{1cm}\\
{\small  V.Bashiry$^1$\thanks {e-mail: bashiry@ipm.ir}\,\,,F.
Falahati$^2$\thanks {e-mail: phy1g832889 @shiraz.ac.ir}\,\,,
} \\
{\small $^1$ Institute for Studies in Theoretical Physics and
Mathematics (IPM),}\\{ \small  P.O. Box 19395-5531, Tehran, Iran }\\
{\small $^2$ Physics Department, Shiraz University, Shiraz, 71454,
Iran }\\}
\date{}
\begin{titlepage}
\maketitle
\thispagestyle{empty}

\begin{abstract}
This study examines the influence of the fourth generation quarks on
the double lepton polarizations forward--backward asymmetries  in $B
\rar K^\ast \ell^+ \ell^-$ decay. We obtain that for both ($\mu, \,
\tau$) channels the magnitude and the sign of the differential
forward--backward asymmetries and the magnitude of the average
forward--backward asymmetries are quite sensitive to the 4th
generation quarks mass and mixing parameters. It can serve as a good
tool to search for new physics effects, precisely, to search for the
fourth generation quarks($t',\, b')$ via its indirect manifestations
in the loop diagrams.
\end{abstract}

~~~PACS numbers: 12.60.--i, 13.30.--a, 14.20.Mr
\end{titlepage}

\section{Introduction}
New Physics (NP) can be researched in two ways: either by raising
the available energy at colliders to produce new particles and
reveal them directly, or by increasing the experimental precision on
certain processes involving Standard Model (SM) particles as
external states. The latter option, indirect search for NP, should
be pursued using processes that are forbidden, and are very rare or
precisely calculable in the SM. In this respect, Flavor Changing
Neutral Current (FCNC) and CP-violating processes are among the most
powerful probes of NP, since in the SM they cannot arise at the
tree-level and even at the loop level they are strongly suppressed
by the GIM mechanism. Furthermore, in the quark sector they are all
calculable in terms of the CKM matrix, and in particular of the
parameters $\bar \rho$ and $\bar \eta$ in the generalized
Wolfenstein parametrization~\cite{Wolfenstein:1983yz}.
Unfortunately, in many cases a deep understanding of hadronic
dynamics is required in order to be able to extract the relevant
short-distance information from measured processes. Lattice QCD and
QCD sum  rules allow us to compute the necessary hadronic parameters
in many processes.  Indeed, the Unitarity Triangle Analysis (UTA)
with Lattice QCD input is extremely successful in determining $\bar
\rho$ and $\bar \eta$ and in constraining NP
contributions~\cite{hep-ph/0501199,hep-ph/0606167,hep-ph/0509219,hep-ph/0605213,hep-ph/0406184}.

Once the CKM matrix is precisely determined by means of the UTA, it
is possible to search for NP contributions. FCNC and CP-violating
are indeed the most sensitive probes of NP contributions to penguin
operators. One of the possible extension of the SM is the standard
model with more than 3 generations. Nothing in the standard model
itself fixes the number of quarks and leptons that can exist.
Therefore, the up/down quarks are first generation quarks, while the
electron and e- neutrino are the first generation leptons. Since the
first three generations are full, any new quarks and leptons would
be members of a "fourth generation." In this sense, SM may be
treated as an effective theory of fundamental interactions rather
than fundamental particles. The Democratic Mass Matrix approach
\cite{harari}, which is quite natural in the SM framework, may be
considered as the interesting step in true direction. It is
intriguing that Flavors Democracy favors the existence of the fourth
SM family \cite{datta, celikel,Sultansoy:2000dm}. Any study related
to the decay of 4th generation quarks or indirect effects of those
in FCNC requires the choice of the quark masses which are not free
parameter, rather they are constrained by the experimental value of
 $\rho$ and $S$ parameters \cite{Sultansoy:2000dm}. The $\rho$
 parameter, in terms of the transverse part of the W--and  Z--boson
 self energies at zero momentum transfer, is given in \cite{Djouadi},
 \bea \rho=\frac{1}{1-\Delta\rho};\,\,\ \Delta\rho=\frac{\Pi_{ZZ}(0)}
 {M_Z^2}-\frac{\Pi_{WW}(0)}{M_W^2},\eea
 the common mass of the fourth quark ($m_{t'}$) lies between
320 GeV and 730 GeV considering the experimental value of
$\rho=1.0002^{+0.0007}_{-0.0004}$ \cite{PDG}. The last value is
close to upper limit on heavy quark masses, $m_q\leq 700$ GeV
$\approx 4m_t$, which follows from partial-wave unitarity at high
energies \cite{chanowitz}. It should be noted that with preferable
value $a\approx g_w$ Flavor Democracy predicts $m_{t'}\approx 8
m_w\approx 640$ GeV. The above mentioned values for mass of $m_{t'}$
disfavors the fifth SM family both because in general we expect that
$m_t\leq m_{t'}\leq m_t^{''}$ and the experimental values of the
$\rho$ and $S$ parameters \cite{Sultansoy:2000dm} restrict the quark
mass up to $700$ Gev.

The study of production, decay channels and LHC signals of the 4th
generation quarks have been continuing. But, one  of the efficient
ways to establish the existence of the 4th generation
 is via their indirect manifestations in loop diagrams. Rare decays, induced
 by flavor changing neutral
current (FCNC) $b \rar s(d)$ transitions are at the forefront of our
quest to understand flavor and the origins of CPV, offering one of
the best probes for New Physics (NP) beyond the Standard Model (SM).
Several hints for NP have emerged in the past few years.
For example, a large difference is seen in direct CP asymmetries in
$B\to K\pi$ decays~\cite{HFAG},
\begin{eqnarray}
{\cal A}_{K\pi}
 \equiv A_{\rm CP}(B^0\to K^+\pi^-) = -0.093 \pm 0.015, &&\nonumber\\
{\cal A}_{K\pi^0}
 \equiv A_{\rm CP}(B^+\to K^+\pi^0) = +0.047 \pm 0.026, &&
 \label{data}
\end{eqnarray}
or $\Delta{\cal A}_{K\pi} \equiv {\cal A}_{K\pi^0}-{\cal A}_{K\pi} =
(14\pm 3)\%$~\cite{Barlow}. As this percentage was not predicted
when first measured in 2004, it has stimulated discussion on the
potential mechanisms that it may have been missed in the SM
calculations \cite{BBNS,KLS,BPS05}.

Better known is the mixing-induced CP asymmetry ${\cal S}_f$
measured in a multitude of CP eigenstates $f$. For penguin-dominated
$b \to sq\bar q$ modes, within SM, ${\cal S}_{sq\bar q}$ should be
close to that extracted from $b\to c\bar cs$ modes. The latter is
now measured rather precisely, ${\cal S}_{c\bar cs}=\sin2\phi_1 =
0.674 \pm 0.026$~\cite{Hazumi}, where $\phi_1$ is the weak phase in
$V_{td}$. However, for the past few years, data seem to indicate, at
2.6$\,\sigma$ significance,
\begin{eqnarray}
\Delta {\cal S} \equiv {\cal S}_{sq\bar q}-{\cal S}_{c\bar cs}\leq
0,
 \label{DelS}
\end{eqnarray}
which has stimulated even more discussions.

The $b \rar s(d) \ell^+ \ell^-$ decays has received considerable
attention as a potential testing ground for the effective
Hamiltonian describing FCNC in B and $\Lambda_b$ decay. This
Hamiltonian contains the one--loop effects of the electroweak
interaction, which are sensitive to the quarks contribute in the
loop \cite{Willey}--\cite{Buras1995}. In addition, there are
important QCD corrections, which have recently been calculated in
the NNLL\cite{NNLL}. Moreover, $b \rar s(d) \ell^+ \ell^-$ decay
is also very sensitive to the new physics beyond SM. New physics
effects manifest themselves in rare decays in two different ways,
either through new combinations to the new Wilson coefficients or
through the new operator structure in the effective Hamiltonian,
which is absent in the SM. A crucial problem in the new physics
search within flavour physics is the optimal separation of new
physics effects from uncertainties. It is well known that
inclusive decay modes are dominated partonic contributions;
non--perturbative corrections are in general rather
small\cite{Hurth}. Also ratios of exclusive decay modes such as
asymmetries for $B\rar K(~K^\ast,~\rho,~\gamma)~ \ell^+ \ell^-$
decay \cite{R4621}--\cite{bashirychin} are well studied for
new--physics search. Here large parts of the hadronic
uncertainties partially cancel out.

In this paper we investigate the possibility of searching for new
physics in the heavy baryon decays $B \rar K^\ast \ell^+ \ell^-$
using the SM with four generations of quarks($b',\, t'$). The fourth
quark ($t'$), like $u,c,t$ quarks, contributes in the $b \rar s(d) $
transition at loop level. It would, Clearly, change the branching
ratio and asymmetries. Note that, fourth generation effects on the
branching ratio have been widely studied in baryonic and
semileptonic B decays \cite{Hou:2006jy}--\cite{Turan:2005pf}. But,
there are few works(lepton polarization asymmetries in
$\Lambda_b\rightarrow\Lambda l^{+}l^{-}$\cite{Bashiry2007,FZVB})
related to the study of asymmetries either in heavy baryon to light
baryon decay or in various B  decay channels.

The main problem for the description of exclusive decays is to
evaluate the form factors, i.e., matrix elements of the effective
Hamiltonian between initial and final hadron states. It is well
known that in order to describe   $B \rar K^\ast \ell^+ \ell^-$
decay a number of form factors are needed (see for example
\cite{Aliev:2004hi}).

It should be mentioned here that the exclusive decay  $ B \rar
K^\ast \ell^+ \ell^-$ decay rate, lepton polarization and CP
asymmetry are studied widely in the SM and beyond the SM i.e.,
\cite{Aliev:2004hi}, \cite{Arhrib:2002md}.

The sensitivity of the CP asymmetry  to the existence of fourth
generation quarks in the $B \rar K^\ast \ell^+ \ell^-$ decay is
investigated in \cite{Arhrib:2002md} and it is obtained that the CP
asymmetry is very sensitive to the fourth generation parameters
($m_{t'}$, $V_{t'b}V^*_{t's}$ ). In this connection it is natural to
ask whether polarized lepton pair forward--backward asymmetries are
sensitive to the fourth generation parameters, in the same decay. In
the present work we try to answer to this question.

The paper is organized as follows. In section 2, using the effective
hamiltonian, the general expressions for the matrix element of $ B
\rar K^\ast \ell^+ \ell^-$  decay is derived. Section 3 devoted to
the calculations of polarized lepton pair forward--backward
asymmetries. In section 4 we investigate the sensitivity of these
functions to the fourth generation parameters ($m_{t'}$, $r_{sb},\,\
\phi_{sb}$ ).

\section{Strategy}
With a sequential fourth generation, the Wilson coefficients $C_7,\,
C_9$ and $C_{10}$ receive contributions from the $t'$ quark loop,
which we will denote as $C^{new}_{7,9,10}$ . Because a sequential
fourth generation couples in a similar way to the photon and W, the
effective Hamiltonian relevant for $b \rar s \ell^+ \ell^-$ decay
has the following form:
 \bea\label{Hgen} {\cal H}_{eff} &=& \frac{4 G_F}{\sqrt{2}}
V_{tb}V_{ts}^\ast \sum_{i=1}^{10} {\cal C}_i(\mu) \, {\cal
O}_i(\mu)~, \eea where the full set of the operators ${\cal
O}_i(\mu)$ and the corresponding expressions for the Wilson
coefficients ${\cal C}_i(\mu)$ in the SM are given in
\cite{R23}--\cite{R24}. As it has already been noted , the fourth
generation up type quark $t'$ is introduced in the same way as $u,\
c,\ t$ quarks introduce in the SM, and so new operators do not
appear and clearly the full operator set is exactly the same as in
SM. The fourth generation changes the values of the Wilson
coefficients $C_7(\mu),~C_9(\mu)$ and $C_{10}(\mu)$, via virtual
exchange of the fourth generation up type quark $t^\prime$. The
above mentioned Wilson coefficients will explicitly change
  as
\bea\lambda_t C_i \rightarrow \lambda_t C^{SM}_i+\lambda_{t'}
C^{new}_i~,\eea where $\lambda_f=V_{f b}^\ast V_{fs}$. The unitarity
of the $4\times4$ CKM matrix leads to
\bea\lambda_u+\lambda_c+\lambda_t+\lambda_{t'}=0.\eea\ Since
$\lambda_u=V_{ub}^\ast V_{us}$ is very small in strength compared to
the others . Then $\lambda_t\approx -\lambda_c-\lambda_{t'}$ and
$\lambda_c=V_{c b}^\ast V_{cs}\approx 0.04$ is real by convention.
It follows that \bea \lambda_t C^{SM}_i+\lambda_{t'}
C^{new}_i=\lambda_c C^{SM}_i+\lambda_{t'} (C^{new}_i-C^{SM}_i )\eea
It is clear that, for the $m_{t'}\rar m_t$ or $\lambda_{t'}\rar 0$,
$\lambda_{t'} (C^{new}_i-C^{SM}_i )$ term vanishes, as required by
the GIM mechanism. One can also write $C_i$'s in the following form
\bea\label{c4} C_7^{tot}(\mu) &=& C_7^{SM}(\mu) +
\frac{\lambda_{t'}}
{\lambda_t} C_7^{new} (\mu)~, \nnb \\
C_9^{tot}(\mu) &=& C_9^{SM}(\mu) +  \frac{\lambda_{t'}}
{\lambda_t}C_9^{new} (\mu) ~, \nnb \\
C_{10}^{tot}(\mu) &=& C_{10}^{SM}(\mu) +  \frac{\lambda_{t'}}
{\lambda_t} C_{10}^{new} (\mu)~, \eea
where the last terms in these expressions describe the contributions
of the $t^\prime$ quark to the Wilson coefficients. $\lambda_{t'}$  can be parameterized
as : \bea {\label{parameter}} \lambda_{t'}=V_{t^\prime b}^\ast V_{t^\prime
s}=r_{sb}e^{i\phi_{sb}}\eea

 In deriving Eq. (\ref{c4}) we factored
out the term $V_{tb}^\ast V_{ts}$ in the effective Hamiltonian given
in Eq. (\ref{Hgen}). The explicit forms of the $C_i^{new}$ can
easily be obtained from the corresponding expression of the Wilson
coefficients in SM by substituting $m_t \rar m_{t^\prime}$ (see
\cite{R23,R25}). If the $s$ quark mass is neglected, the above
effective Hamiltonian leads to following matrix element for the $b
\rar s \ell^+ \ell^-$ decay \bea\label{e1} {\cal H}_{eff} &=&
\frac{G\alpha}{2\sqrt{2} \pi}
 V_{tb}V_{ts}^\ast
\Bigg[ C_9^{tot} \, \bar s \gamma_\mu (1-\gamma_5) b \, \bar \ell
\gamma_\mu \ell + C_{10}^{tot} \bar s \gamma_\mu (1-\gamma_5) b \,
\bar \ell \gamma_\mu \gamma_5 \ell \nnb \\
&-& 2  C_7^{tot}\frac{m_b}{q^2} \bar s \sigma_{\mu\nu} q^\nu
(1+\gamma_5) b \, \bar \ell \gamma_\mu \ell \Bigg]~, \eea where
$q^2=(p_++p_-)^2$ and $p_-$ and $p_+$ are the final leptons
four--momenta. The effective coefficient $C_9^{tot}$ can be
written in the following form \bea C_9^{tot} = C_9 + Y(s')~, \eea
where $s' = q^2 / m_b^2$ and the function $Y(s')$ denotes the
perturbative part coming from one loop matrix elements of four
quark operators and is given\cite{R23,R24}, \bea Y_{per}(s') &=&
g(\hat m_c, s') (3 C_1 + C_2 + 3 C_3 + C_4 + 3 C_5 + C_6) \nnb
\\&-& \frac{1}{2} g(1, s') (4 C_3 + 4 C_4 + 3 C_5 + C_6) \nnb
\\&-& \frac{1}{2} g(0, s') (C_3 + 3 C_4) + \frac{2}{9} (3 C_3 +
C_4 + 3 C_5 + C_6)~, \eea where $\hat m_c = \frac{m_c}{m_b} $. The
explicit expressions for $g(\hat m_c, s')$, $g(0, s')$, $g(1, s')$
and the values of $C_i$ in the SM can be found in Table 1
\cite{R23,R24}.
\begin{table}
\renewcommand{\arraystretch}{1.5}
\addtolength{\arraycolsep}{3pt}
$$
\begin{array}{|c|c|c|c|c|c|c|c|c|}
\hline C_{1} & C_{2} & C_{3} & C_{4} & C_{5} & C_{6} & C_{7}^{SM} &
C_{9}^{SM} & C_{10}^{SM}\\ \hline
-0.248 & 1.107& 0.011& -0.026& 0.007& -0.031& -0.313& 4.344& -4.669\\
\hline
\end{array}
$$
\caption{The numerical values of the Wilson coefficients at $\mu =
m_{b}$ scale within the SM. The corresponding numerical value of
$C^{0}$ is $0.362$.}
\renewcommand{\arraystretch}{1}
\addtolength{\arraycolsep}{-3pt}
\end{table}

In addition to the short distance contribution, $Y_{per}(s')$
receives also long distance contributions, which have their origin
in the real $c\bar c$ intermediate states, i.e., $J/\psi$,
$\psi^\prime$, $\cdots$. The $J/\psi$ family is introduced by the
Breit--Wigner distribution for the resonances through the
replacement \cite{R26}--\cite{R28} \bea Y(s') = Y_{per}(s') +
\frac{3\pi}{\alpha^2} \, C^{(0)} \sum_{V_i=\psi_i} \kappa_i \,
\frac{m_{V_i} \Gamma(V_i \rar \ell^+ \ell^-)} {m_{V_i}^2 - s' m_b^2
- i m_{V_i} \Gamma_{V_i}}~, \eea where $C^{(0)}= 3 C_1 + C_2 + 3 C_3
+ C_4 + 3 C_5 + C_6$. The phenomenological parameters $\kappa_i$ can
be fixed from ${\cal B} (B \rar K^\ast V_i \rar K^\ast \ell^+
\ell^-) = {\cal B} (B \rar K^\ast V_i)\, {\cal B} ( V_i \rar \ell^+
\ell^-)$, where the data for the right hand side is given in
\cite{R29}. For the lowest resonances $J/\psi$ and $\psi^\prime$ one can
use $\kappa = 1.65$ and $\kappa = 2.36$, respectively (see \cite{R30}).

After having an idea of the effective Hamiltonian and the relevant
Wilson coefficients, we now proceed to evaluate the transition
matrix elements for the process $B (p_{B}) \to K^\ast
(p_{K^\ast})~ l^+ (p_+)l^-(p_-)$. It follows from Eq. (\ref{e1})
that in order to calculate the decay width and other physical
observables of the exclusive $B \rar K^\ast \ell^+ \ell^-$ decay,
the matrix elements $\lla K^\ast \vel \bar s \gamma_\mu (1 -
\gamma_5) b \ver B \rra$ and $\lla K^\ast \vel \bar s i
\sigma_{\mu\nu} q^\nu (1 + \gamma_5) b \ver \rra$ have to be
calculated. In other words, the exclusive $B \rar K^\ast \ell^+
\ell^-$ decay which is described in terms of the matrix elements
of the quark operators given in Eq. (\ref{e1}) over meson states,
can be parameterized in terms of form factors. For the vector
meson $K^\ast$ with polarization vector $\varepsilon_\mu$ the
semileptonic form factors of the V--A current is defined as \bea
\lefteqn{ \lla K^\ast(p,\varepsilon) \vel \bar s \gamma_\mu (1 -
\gamma_5) b \ver
B(p_B) \rra =} \nnb \\
&&- \epsilon_{\mu\nu\rho\sigma} \varepsilon^{\ast\nu} p^\rho
q^\sigma \frac{2 V(q^2)}{m_B+m_{K^\ast}} - i \varepsilon_\mu
(m_B+m_{K^\ast}) A_1(q^2) + i (p_B + p_{K^\ast})_\mu
(\varepsilon^\ast q)
\frac{A_2(q^2)}{m_B+m_{K^\ast}} \nnb \\
&&+ i q_\mu \frac{2 m_{K^\ast}}{q^2} (\varepsilon^\ast q)
\left[A_3(q^2)-A_0(q^2)\right]~, \eea where $\varepsilon$ is the
polarization vector of $K^\ast$ meson and $q=p_B-p_{K^\ast}$ is the
momentum transfer. Using the equation of motion, the form factor
$A_3(q^2)$ can be written in terms of the form factors $A_1(q^2)$,
$A_2(q^2)$ as follows \bea A_3 = \frac{m_B+m_{K^\ast}}{2 m_{K^\ast}}
A_1 - \frac{m_B-m_{K^\ast}}{2 m_{K^\ast}} A_2~. \eea In order to
ensure finiteness of (8) at $q^2=0$, we demand that $A_3(q^2=0) =
A_0(q^2=0)$. The semileptonic form factors coming from the dipole
operator $\sigma_{\mu\nu} q^\nu (1 + \gamma_5) b$ are defined as
\bea \lefteqn{ \lla K^\ast(p,\varepsilon) \vel \bar s i
\sigma_{\mu\nu} q^\nu (1 + \gamma_5) b \ver
B(p_B) \rra =} \nnb \\
&&4 \epsilon_{\mu\nu\rho\sigma} \varepsilon^{\ast\nu} p^\rho
q^\sigma T_1(q^2) + 2 i \left[ \varepsilon_\mu^\ast
(m_B^2-m_{K^\ast}^2) -
(p_B + p_{K^\ast})_\mu (\varepsilon^\ast q) \right] T_2(q^2) \nnb \\
&&+ 2 i (\varepsilon^\ast q) \left[ q_\mu - (p_B + p_{K^\ast})_\mu
\frac{q^2}{m_B^2-m_{K^\ast}^2} \right] T_3(q^2)~. \eea From Eqs.
(7), (9) and (10) we observe that in calculating the physical
observable at hadronic level, i.e., for the $B \rar K^\ast \ell^+
\ell^-$ decay, we face the problem of computing the form factors.
This problem is related to the nonperturbative sector of QCD and it
can be solved only in framework a nonperturbative approach. In the
present work we choose light cone QCD sum rules method predictions
for the form factors. In what follows we will use the results of the
work \cite{R31,R32,R33} in which the form factors are described by a
three--parameter fit where the radiative corrections up to leading
twist contribution and SU(3)--breaking effects are taken into
account. And leading to  \bea F(q^{2})\in\{V(q^2), A_{0}(q^{2}),
B_0(q^{2}), A_{2}(q^{2}), A_{3}(q^{2}), T_{1}(q^{2}), T_{2}(q^{2}),
T_{3}(q^{2})\}~,\nnb \eea the $q^{2}$--dependence of any of these
form factors could be parameterized as \cite{R31,R32} \bea
\label{formfac} F(s) = \frac{F(0)}{1-a_F\,s + b_F\, s^{2}}~, \eea
where the parameters $F(0)$, $a_F$ and $b_F$ are listed in Table 3
for each form factor.
\begin{table}[h]
\renewcommand{\arraystretch}{1.5}
\addtolength{\arraycolsep}{3pt}
$$
\begin{array}{|l|ccc|}
\hline & F(0) & a_F & b_F \\ \hline
A_0^{B \rar K^*} &\phantom{-}0.47 & 1.64 & \phantom{-} 0.94 \\
A_1^{B \rar K^*} &\phantom{-}0.35 & 0.54 & -0.02 \\
A_2^{B \rar K^*} &\phantom{-}0.30 & 1.02 & \phantom{-} 0.08\\
V^{B \rar K^*}   &\phantom{-}0.47 & 1.50 & \phantom{-} 0.51\\
T_1^{B \rar K^*} &\phantom{-}0.19 & 1.53 & \phantom{-} 1.77\\
T_2^{B \rar K^*} &\phantom{-}0.19 & 0.36 & -0.49\\
T_3^{B \rar K^*} &\phantom{-}0.13 & 1.07 & \phantom{-} 0.16\\ \hline
\end{array}
$$
\caption{The form factors for $B\rightarrow K^\ast \ell^{+}\ell^{-}$
in a three--parameter fit \cite{R31}.}
\renewcommand{\arraystretch}{1}
\addtolength{\arraycolsep}{-3pt}
\end{table}

Using the form factors, the matrix element of the $B \rar K^\ast
\ell^+ \ell^-$ decay takes the following form \bea \lefteqn{
\label{e6306} {\cal M}(B\rightarrow K^\ast \ell^{+}\ell^{-}) =
\frac{G \alpha}{4 \sqrt{2} \pi} V_{tb} V_{ts}^\ast }\\
&&\times \Bigg\{ \bar \ell \gamma^\mu(1-\gamma_5) \ell \, \Big[ -2
B_0 \epsilon_{\mu\nu\lambda\sigma} \varepsilon^{\ast\nu}
p_{K^\ast}^\lambda q^\sigma
 -i B_1 \varepsilon_\mu^\ast
+ i B_2 (\varepsilon^\ast q) (p_B+p_{K^\ast})_\mu
+ i B_3 (\varepsilon^\ast q) q_\mu  \Big] \nnb \\
&&+ \bar \ell \gamma^\mu(1+\gamma_5) \ell \, \Big[ -2 C_1
\epsilon_{\mu\nu\lambda\sigma} \varepsilon^{\ast\nu}
p_{K^\ast}^\lambda q^\sigma
 -i D_1 \varepsilon_\mu^\ast
+ i D_2 (\varepsilon^\ast q) (p_B+p_{K^\ast})_\mu + i D_3
(\varepsilon^\ast q) q_\mu  \Big] \Bigg\}~,\nnb \eea where \bea
\label{e6307} B_0 &=& (C_9^{tot}- C_{10}^{tot})
\frac{V}{m_B+m_{K^\ast}} +
4 (m_b+m_s)C_7^{tot} \frac{T_1}{q^2} ~, \nnb \\
B_1 &=& (C_9^{tot}- C_{10}^{tot}) (m_B+m_{K^\ast}) A_1 + 4
(m_b-m_s)C_7^{tot} (m_B^2-m_{K^\ast}^2)
\frac{T_2}{q^2} ~, \nnb \\
B_2 &=& \frac{C_9^{tot}- C_{10}^{tot}}{m_B+m_{K^\ast}} A_2 + 4
(m_b-m_s)C_7^{tot} \frac{1}{q^2}  \left[ T_2 +
\frac{q^2}{m_B^2-m_{K^\ast}^2} T_3 \right]~,
\nnb \\
B_3 &=& 2 (C_9^{tot}- C_{10}^{tot}) m_{K^\ast} \frac{A_3-A_0}{q^2}-4
 (m_b-m_s)C_7^{tot} \frac{T_3}{q^2} ~, \nnb \\
C_1 &=& B_0 ( C_{10}^{tot} \rar -C_{10}^{tot}\nnb)~, \\
D_i &=& B_i ( C_{10}^{tot} \rar -C_{10}^{tot}\nnb)~,~~~~(i=1,~2,~3).
\eea

From this expression of the decay amplitude, for the differential
decay width we get the following result: \bea \label{e6308}
\frac{d\Gamma}{d\hat{s}}(B \rar K^\ast \ell^+ \ell^-) = \frac{G^2
\alpha^2 m_B}{2^{14} \pi^5} \vel V_{tb}V_{ts}^\ast \ver^2
\lambda^{1/2}(1,\hat{r},\hat{s}) v \Delta(\hat{s})~, \eea with
\bea
\label{e6309} \Delta&=&\frac{2}{3\hat{r}_{K^{*}}\hat{s}}m_{B}^2Re[
-12m_{B}^2\hat{m_{l}}^2\lambda\hat{s}\{(B_{3}-D_{2}-D_{3})B_{1}^{*}-(B_{3}+B_{2}-D_{3})D_{1}^{*}\}\nnb\\
\nonumber&+&12m_{B}^4\hat{m_{l}}^2\lambda\hat{s}(1-\hat{r}_{K^{*}})(B_{2}-D_{2})(B_{3}^{*}-D_{3}^{*})\\
\nonumber&+&48\hat{m_{l}}^2\hat{r}_{K^{*}}\hat{s}(3B_{1}D_{1}^{*}+2m_{B}^4\lambda
B_0C_{1}^{*})\\
\nonumber&-&16m_{B}^4\hat{r}_{K^{*}}\hat{s}\lambda(\hat{m_{l}}^2-\hat{s})\{|B_0|^{2}+|C_{1}|^{2}\}\\
\nonumber&-&6m_{B}^4\hat{m_{l}}^2\lambda\hat{s}\{2(2+2\hat{r}_{K^{*}}-\hat{s})B_{2}D_{2}^{*}-\hat{s}|(B_{3}-D_{3})|^{2}\}\\
\nonumber&-&4m_{B}^2\lambda\{\hat{m_{l}}^2(2-2\hat{r}_{K^{*}}+\hat{s})+\hat{s}(1-\hat{r}_{K^{*}}-\hat{s})\}(B_{1}B_{2}^{*}+D_{1}D_{2}^{*})\\
\nonumber&+&\hat{s}\{6\hat{r}_{K^{*}}\hat{s}(3+v^2)+\lambda(3-v^2)\}\{|B_{1}|^{2}+|D_{1}|^{2}\}\\
&-&2m_{B}^4\lambda\{\hat{m_{l}}^2[\lambda-3(1-\hat{r}_{K^{*}})^2]-\lambda\hat{s}\}\{|B_{2}|^{2}+|D_{2}|^{2}\}],\\
\nonumber\eea where $\hat{s}=q^2/m_B^2$,
$\hat{r}_{K^{*}}=m_{K^\ast}^2/m_B^2$ and
$\lambda(a,b,c)=a^2+b^2+c^2-2ab-2ac-2bc$, $\hat{m}_\ell=m_\ell/m_B$,
$v=\sqrt{1-4\hat{m}_\ell^2/\hat{s}}$ is the final lepton velocity.

The definition of the polarized $FB$ asymmetries will be presented
in the next section.

\section{Polarized forward--backward asymmetries of leptons}

In order to now calculate the polarization asymmetries of both
leptons defined in the effective four fermion interaction of Eq.
(\ref{e6306}), we must first define the orthogonal vectors $S$ in
the rest frame of $\ell^-$ and $W$ in the rest frame of $\ell^+$
(where these vectors are the polarization vectors of the leptons).
Note that we shall use the subscripts $L$, $N$ and $T$ to correspond
to the leptons being polarized along the longitudinal, normal and
transverse directions respectively
\cite{Kruger:1996cv,Bensalem:2002ni}.
\begin{eqnarray}
S^\mu_L & \equiv & (0, \mathbf{e}_{L}) ~=~ \left(0,
\frac{\mathbf{p}_-}{|\mathbf{p}_-|}
\right) , \nonumber \\
S^\mu_N & \equiv & (0, \mathbf{e}_{N}) ~=~ \left(0,
\frac{\mathbf{p_{K^\ast}} \times
\mathbf{p}_-}{|\mathbf{p_{K^\ast}} \times
\mathbf{p}_- |}\right) , \nonumber \\
S^\mu_T & \equiv & (0, \mathbf{e}_{T}) ~=~ \left(0, \mathbf{e}_{N}
\times \mathbf{e}_{L}\right) , \label{sec3:eq:1} \\
W^\mu_L & \equiv & (0, \mathbf{w}_{L}) ~=~ \left(0,
\frac{\mathbf{p}_+}{|\mathbf{p}_+|} \right) , \nonumber \\
W^\mu_N & \equiv & (0, \mathbf{w}_{N}) ~=~ \left(0,
\frac{\mathbf{p_{K^\ast}} \times
\mathbf{p}_+}{|\mathbf{p_{K^\ast}} \times
\mathbf{p}_+ |} \right) , \nonumber \\
W^\mu_T & \equiv & (0, \mathbf{w}_{T}) ~=~ (0, \mathbf{w}_{N} \times
\mathbf{w}_{L}) , \label{sec3:eq:2}
\end{eqnarray}
where $\mathbf{p}_+$, $\mathbf{p}_-$ and $\mathbf{p_{K^\ast}}$ are
the three momenta of the $\ell^+$, $\ell^-$ and $K^\ast$ particles
respectively. On boosting the vectors defined by
Eqs.(\ref{sec3:eq:1},\ref{sec3:eq:2}) to the c.m. frame of the
$\ell^- \ell^+$ system only the longitudinal vector will be
boosted, whilst the other two vectors remain unchanged. The
longitudinal vectors after the boost will become;
\begin{eqnarray}
S^\mu_L & = & \left( \frac{|\mathbf{p}_-|}{m_\ell}, \frac{E_{\ell}
\mathbf{p}_-}{m_\ell |\mathbf{p}_-|} \right) , \nonumber \\
W^\mu_L & = & \left( \frac{|\mathbf{p}_-|}{m_\ell}, - \frac{E_{\ell}
\mathbf{p}_-}{m_\ell |\mathbf{p}_-|} \right) . \label{sec3:eq:3}
\end{eqnarray}
The polarization asymmetries can now be calculated using the spin
projector ${1 \over 2}(1 + \gamma_5 \!\!\not\!\! S)$ for $\ell^-$
and the spin projector ${1 \over 2}(1 + \gamma_5\! \not\!\! W)$ for
$\ell^+$.

\par Equipped with the above expressions, we now define the various
forward--backward asymmetries of leptons. The definition of the
unpolarized and normalized differential forward--backward asymmetry
is (see for example \cite{R6322}) \bea \label{e6312} {\cal A}_{FB} =
\frac{\ds \int_{0}^{1} \frac{d^2\Gamma}{d\hat{s} dz} - \int_{-1}^{0}
\frac{d^2\Gamma}{d\hat{s} dz}} {\ds \int_{0}^{1}
\frac{d^2\Gamma}{d\hat{s} dz} + \int_{-1}^{0}
\frac{d^2\Gamma}{d\hat{s} dz}}~, \eea where $z=\cos\theta$ is the
angle between $B$ meson and $\ell^-$ in the center of mass frame of
leptons. When the spins of both leptons are taken into account, the
${\cal A}_{FB}$ will be a function of the spins of the final leptons
and it is defined as: \bea \label{e6313} {\cal A}_{FB}^{ij}(\hat{s})
\es \Bigg(\frac{d\Gamma(\hat{s})}{d\hat{s}} \Bigg)^{-1} \Bigg\{
\int_0^1 dz - \int_{-1}^0 dz \Bigg\} \Bigg\{ \Bigg[
\frac{d^2\Gamma(\hat{s},\vec{s}^{\,-} = \vec{i},\vec{s}^{\,+} =
\vec{j})} {d\hat{s} dz} - \frac{d^2\Gamma(\hat{s},\vec{s}^{\,-} =
\vec{i},\vec{s}^{\,+} = -\vec{j})} {d\hat{s} dz}
\Bigg] \nnb \\
\ek \Bigg[ \frac{d^2\Gamma(\hat{s},\vec{s}^{\,-} =
-\vec{i},\vec{s}^{\,+} = \vec{j})} {d\hat{s} dz} -
\frac{d^2\Gamma(\hat{s},\vec{s}^{\,-} = -\vec{i},\vec{s}^{\,+} =
-\vec{j})} {d\hat{s} dz} \Bigg]
\Bigg\}~,\nnb \\ \nnb \\
\es {\cal A}_{FB}(\vec{s}^{\,-}=\vec{i},\vec{s}^{\,+}=\vec{j})   -
{\cal A}_{FB}(\vec{s}^{\,-}=\vec{i},\vec{s}^{\,+}=-\vec{j})  -
{\cal A}_{FB}(\vec{s}^{\,-}=-\vec{i},\vec{s}^{\,+}=\vec{j})  \nnb \\
\ar {\cal A}_{FB}(\vec{s}^{\,-}=-\vec{i},\vec{s}^{\,+}=-\vec{j})~.
\eea

where the sub-indices $i$ and $j$ can be either $L$, $N$ or $T$.
Using these definitions for the double polarized $FB$ asymmetries,
we get the following results: \begin{eqnarray}
A^{LL}_{FB}&=&\frac{2}{\hat{r}_{K^\ast}\Delta}m_{B}^3\sqrt{\lambda}vRe[
8m_{B}\hat{r}_{K^\ast}\hat{s}(B_0B_{1}^{*}-C_{1}D_{1}^{*})],\\
\nonumber
A^{LN}_{FB}&=&\frac{8}{3\hat{r}_{K^\ast}\Delta\hat{s}}m_{B}^2\sqrt{\hat{s}}\lambda
v Im[-\hat{m_{l}}(B_{1}D_{1}^{*}+m_{B}^4\lambda
B_{2}D_{2}^{*})+4m_{B}^4\hat{m_{l}}\hat{r}_{K^\ast}\sqrt{\hat{s}}B_0C_{1}^{*}\\
&+&m_{B}^2\hat{m_{l}}(1-\hat{r}_{K^\ast}-\hat{s})(B_{1}D_{2}^{*}+B_{2}D_{1}^{*})] ,\\
\nonumber
A^{NL}_{FB}&=&\frac{8}{3\hat{r}_{K^\ast}\Delta\hat{s}}m_{B}^2\sqrt{\hat{s}}\lambda
v Im[-\hat{m_{l}}(B_{1}D_{1}^{*}+m_{B}^4\lambda
B_{2}D_{2}^{*})+4m_{B}^4\hat{m_{l}}\hat{r}_{K^\ast}\sqrt{\hat{s}}B_{0}C_{1}^{*}\\
&+&m_{B}^2\hat{m_{l}}(1-\hat{r}_{K^\ast}-\hat{s})(B_{1}D_{2}^{*}+B_{2}D_{1}^{*})] ,\\
\nonumber
A^{LT}_{FB}&=&\frac{4}{3\hat{r}_{K^\ast}\Delta\hat{s}}m_{B}^2\sqrt{\hat{s}}\lambda
Re[-\hat{m_{l}}\{|B_{1}+D_{1}|^{2}+m_{B}^4\lambda|B_{2}+D_{2}|^{2}\}\\
\nonumber&+&4m_{B}^4\hat{m_{l}}\hat{s}\hat{r}_{K^\ast}\{|B_{0}+C_{1}|^{2}\}\\
&+&2m_{B}^2\hat{m_{l}}(1-\hat{r}_{K^\ast}-\hat{s})(B_{1}+D_{1})(B_{2}^{*}+D_{2}^{*})],\\
\nonumber
A^{TL}_{FB}&=&\frac{4}{3\hat{r}_{K^\ast}\Delta\hat{s}}m_{B}^2\sqrt{\hat{s}}\lambda
Re[\hat{m_{l}}\{|B_{1}+D_{1}|^{2}+m_{B}^4\lambda|B_{2}+D_{2}|^{2}\}\\
\nonumber&-&4m_{B}^4\hat{m_{l}}\hat{s}\hat{r}_{K^\ast}\{|B_{0}+C_{1}|^{2}\}\\
&-&2m_{B}^2\hat{m_{l}}(1-\hat{r}_{K^\ast}-\hat{s})(B_{1}+D_{1})(B_{2}^{*}+D_{2}^{*})],\\
\nonumber
A^{NT}_{FB}&=&\frac{2}{\hat{r}_{K^\ast}\Delta\hat{s}}m_{B}^2\sqrt{\lambda}Im[
-2m_{B}^4\hat{m_{l}}^2\hat{s}\lambda(B_{2}+D_{2})(B_{3}^{*}-D_{3}^{*})\\
\nonumber&+&4m_{B}^4\hat{m_{l}}^2\lambda(1-\hat{r}_{K^\ast})B_{2}D_{2}^{*}\\
\nonumber&+&2m_{B}^2\hat{m_{l}}^2\hat{s}(1+3\hat{r}_{K^\ast}-\hat{s})(B_{1}B_{2}^{*}-D_{1}D_{2}^{*})\\
\nonumber&+&\hat{m_{l}}(1-\hat{r}_{K^\ast}-\hat{s})\{+2\hat{s}m_{B}^{2}\hat{m_{l}}(B_{1}+D_{1})(B_{3}^{*}-D_{3}^{*})\\
\nonumber&+&4\hat{m_{l}}B_{1}D_{1}^{*}\}\\
&+&2m_{B}^2\hat{m_{l}}^2[\lambda+(1-\hat{r}_{K^\ast}-\hat{s})(1-\hat{r}_{K^\ast})](B_{1}^{*}D_{2}+B_{2}^{*}D_{1})],\\
\nonumber
A^{TN}_{FB}&=&\frac{2}{\hat{r}_{K^\ast}\Delta\hat{s}}m_{B}^2\sqrt{\lambda}Im[
-2m_{B}^4\hat{m_{l}}^2\hat{s}\lambda(B_{2}+D_{2})(B_{3}^{*}-D_{3}^{*})\\
\nonumber&+&4m_{B}^4\hat{m_{l}}^2\lambda(1-\hat{r}_{K^\ast})B_{2}D_{2}^{*}\\
\nonumber&+&2m_{B}^2\hat{m_{l}}^2\hat{s}(1+3\hat{r}_{K^\ast}-\hat{s})(B_{1}B_{2}^{*}-D_{1}D_{2}^{*})\\
\nonumber&+&\hat{m_{l}}(1-\hat{r}_{K^\ast}-\hat{s})\{
-2\hat{s}m_{B}^{2}\hat{m_{l}}(B_{1}+D_{1})(B_{3}^{*}-D_{3}^{*})\\
\nonumber&+&4\hat{m_{l}}B_{1}D_{1}^{*}\}\\
&+&2m_{B}^2\hat{m_{l}}^2[\lambda+(1-\hat{r}_{K^\ast}-\hat{s})(1-\hat{r}_{K^\ast})](B_{1}^{*}D_{2}+B_{2}^{*}D_{1})],\\
A^{NN}_{FB}&=&0\\
A^{TT}_{FB}&=&0\\
\nonumber
\end{eqnarray}

\section{Numerical analysis}

In this part we examine the dependence the polarized $FB$ asymmetry
to the fourth quark mass($m_{t'}$) and the product of quark mixing
matrix elements ($V_{t^\prime b}^\ast V_{t^\prime
s}=r_{sb}e^{i\phi_{sb}}$). The input parameters we use in our
numerical calculations are: $\vel V_{tb} V_{ts}^\ast \ver =
0.0385,~m_{K^\ast}=0.892~GeV,~~m_{\tau}=1.77~GeV,~~m_{\mu}=0.106~GeV,~
~m_{b}=4.8~GeV,~~m_{B}=5.26~GeV$ and $\Gamma_B = 4.22\times
10^{-13}~GeV$. For the values of the Wilson coefficients we use
$C_7^{SM}=-0.313,~C_9^{SM}=4.344$ and $C_{10}^{SM}=-4.669$. It
should be noted that the above--presented value for $C_9^{SM}$
corresponds only to short distance contributions. In addition to the
short distance contributions, it receives long distance
contributions which result from the conversion of $\bar{c}c$ to the
lepton pair. In this work we neglect long distance contributions.
The reason for such a choice is dictated by the fact that, in the SM
the zero position of ${\cal A}_{FB}$ for the $B \rar K^\ast \ell^+
\ell^-$ decay is practically independent of the form factors and is
determined in terms of short distance Wilson coefficients $C_9^{SM}$
and $C_7^{SM}$ (see \cite{R30, R6312}) and $s_0=3.9~GeV^2$. For the
form factors we have used the light cone QCD sum rules results
\cite{R6324,R6325}. As a result of the analysis carried out in this
scheme, the $q^2$ dependence of the form factors can be represented
in terms of three parameters as (\ref{formfac}),  where the values
of parameters $F(0)$, $a_F$ and $b_F$ for the $B \rar K^\ast$ decay
are listed in Table 2.

 In order to perform quantitative analysis of the
polarized $FB$ asymmetries  the values of the new
parameters($m_{t'},\,r_{sb},\,\phi_{sb}$) are needed. Using the
experimental values of $B\rar X_s \gamma$ and $B\rar X_s \ell^+
\ell^-$, the bound on $r_{sb}\sim\{0.01-0.03\}$ has been obtained
\cite{Arhrib:2002md} for $\phi_{sb}\sim\{0-2\pi\}$ and
$m_{t'}\sim\{300,400\}~$(GeV). We do the complete analysis about the
range of the new parameters considering the recent experimental
value of the ${\cal{B}}_r (B\rar X_s \ell^+
\ell^-=(1.59\pm0.5)\times 10^{-6})$\cite{HFAG}. Now, we have
obtained that in the case of the $1\sigma$ level deviation from the
measured branching ratio the maximum values of $m_{t'}$ are below
than the theoretical upper limits. The results are shown in Table 3,
4 and 5 \cite{FZVB}.
\begin{table}
\renewcommand{\arraystretch}{1.5}
\addtolength{\arraycolsep}{3pt}
$$
\begin{array}{|c|c|c|c |c|}
\hline  r_{sb} & 0.005 & 0.01 &0.02 & 0.03 \\
\hline
m_{t'}(GeV) & 739 &529 & 385 & 331\\
\hline
\end{array}
$$
\caption{The  experimental limit on the maximum value of $ m_{t'}
$ for $\phi_{sb}=\pi/3$}
\renewcommand{\arraystretch}{1}
\addtolength{\arraycolsep}{-3pt}
\end{table}
\begin{table}
\renewcommand{\arraystretch}{1.5}
\addtolength{\arraycolsep}{3pt}
$$
\begin{array}{|c|c|c|c |c|}
\hline  r_{sb} & 0.005 & 0.01 &0.02 & 0.03 \\
\hline
m_{t'}(GeV) &511 &373 & 289 & 253\\
\hline
\end{array}
$$
\caption{The  experimental limit on the maximum value of $ m_{t'}
$ for  $\phi_{sb}=\pi/2$}
\renewcommand{\arraystretch}{1}
\addtolength{\arraycolsep}{-3pt}
\end{table}
\begin{table}
\renewcommand{\arraystretch}{1.5}
\addtolength{\arraycolsep}{3pt}
$$
\begin{array}{|c|c|c|c |c|}
\hline  r_{sb} & 0.005 & 0.01 &0.02 & 0.03 \\
\hline
m_{t'}(GeV) &361 &283 & 235 & 217\\
\hline
\end{array}
$$
\caption{The  experimental limit on the maximum value of $ m_{t'}
$ for $\phi_{sb}=2\pi/3$}
\renewcommand{\arraystretch}{1}
\addtolength{\arraycolsep}{-3pt}
\end{table}

In the foregoing numerical analysis, we vary $m_{t'}$ in the range
$175\le m_{t'} \le 600$GeV. The lower range is because of the fact
that the fourth generation up quark should be heavier than the third
ones($m_t \leq m_{t'}$)\cite{Sultansoy:2000dm}. The upper range
comes from the experimental bounds on the $\rho$ and $S$ parameters
of SM, which we mentioned above(see Introduction).
\begin{itemize}
\item In Figs. 1, 2 and 3 we present the dependence of the ${\cal
A}_{FB}^{LL}$ on $q^2$ for the $B\rar K^\ast \mu^+ \mu^-$ at four
fixed values of $m_{t'}:300, 400, 500, 600$ GeV,
$\phi_{sb}=60^\circ$ and $r_{sb}:0.01, 0.02, 0.03$, respectively.
From these figures, we see that the above mentioned
 values of the SM4 parameters slightly shift the zero position of ${\cal A}_{FB}^{LL}$
corresponding to the SM3 result. In other words, our analysis shows
that the zero position of ${\cal A}_{FB}^{LL}$ for the $B\rar K^\ast
\mu^+ \mu^-$ decay is practically independent of the existence of
such SM4 parameters.The magnitude of the ${\cal A}_{FB}^{LL}$ is the
suppression function of SM4 parameters, the greater the values of
$m_{t'}$ and $r_{sb}$ are, the smaller the magnitude of the  ${\cal
A}_{FB}^{LL}$ is. The same situation holds for $B\rar K^\ast \tau^+
\tau^-$ decay(see figs 4, 5 and 6). But, in the case of $B\rar
K^\ast \tau^+ \tau^-$ decay, the zero position for the double
polarization asymmetries ${\cal A}_{FB}^{LL}$ is absent both in SM3
and SM4(see figs 4, 5, 6, 10, 11, 12, 16, 17 and 18).

\item Figs. 7, 8 and 9  depict the dependence of ${\cal
A}_{FB}^{LL}$ on $q^2$ for the $B\rar K^\ast \mu^+ \mu^-$ at four
fixed values of $m_{t'}:300, 400, 500, 600$ GeV,
$\phi_{sb}=90^\circ$ and $r_{sb}:0.01, 0.02, 0.03$, respectively. We
observe from these figures that the zero positions of  ${\cal
A}_{FB}^{LL}$ are much more sensitive to the existence of new quarks
doublet than the case of $\phi=60^\circ$. More precisely, the zero
positions of  ${\cal A}_{FB}^{LL}$ shift to the right of the SM3
point($\Delta q^2=q_{Max}^2-q_{SM}^2\approx 1.5$GeV$^2$ in the best
case)(see Fig. 9). Although, ${\cal A}_{FB}^{LL}$ is suppressed by
some values of $m_{t'}$. However, for some other values of
$m_{t'}$(i.e., $m_{t'}=500)$ the variation of ${\cal A}_{FB}^{LL}$
is opposite to the SM3 case. Therefore, if the sign of the polarized
${\cal A}_{FB}^{LL}$ asymmetries are measured in the experiments in
future, these results are unambiguous indication of the existence of
new physics beyond the SM, more specifically, the existence of
fourth generation of quarks. The same situation happens for $B\rar
K^\ast \tau^+ \tau^-$ decay(see figs 10, 11 and 12).

Much more discussions about Figs. 7--12 are seen as follows:

{\bf I.} If a measurement shows us that the sign is opposite to the
SM3 case, there are two consequences: Firstly, it will obviously
indicate that $\phi_{sb}=90^\circ$. Secondly, it will indicate the
lower limit on the fourth generation up type quark mass. It should
be noted that the large mass of $m_{t'}$(i.e., the mass in the order
of 500GeV) excluded in the $1\sigma$ level deviation of the
branching ratio of $B\rar X_s \ell^+ \ell^-$(see Table 3, 4, 5).
But, the large mass can be included regarding the higher order
deviations(i.e., $2\,,3\sigma$ level).

 {\bf II.} If a measurement of  the sign is agree with the SM3 case, then, similar to the
experimental limits come from the measured branching ratio of the
$B\rar X_s \ell^+ \ell^-$(see Table 3, 4, 5) it will put upper limit
on  the mass of $m_{t'}$.

\item In Figs. 13, 14 and 15 we present the dependence of the ${\cal
A}_{FB}^{LL}$ on $q^2$ for the $B\rar K^\ast \mu^+ \mu^-$ at four
fixed values of $m_{t'}:300, 400, 500, 600$ GeV,
$\phi_{sb}=120^\circ$ and $r_{sb}:0.01, 0.02, 0.03$, respectively.
From these figures we see that the general variation of ${\cal
A}_{FB}^{LL}$ is similar to the case of $\phi_{sb}=60^\circ$.
However, both the the zero position of ${\cal A}_{FB}^{LL}$ and the
magnitude of the ${\cal A}_{FB}^{LL}$ show weaker dependency than
the $\phi_{sb}=60^\circ$ case. The same situation is for $B\rar
K^\ast \tau^+ \tau^-$ decay(see figs 16, 17 and 18).

\item Except ${\cal A}_{FB}^{LL}$, the values of ${\cal A}_{FB}^{ij}$
are quite small for the other components both in the SM3 and the
SM4, whose measurement in the experiments could practically be
impossible. For this reason we do not present the dependencies of
${\cal A}_{FB}^{ij}$ on $q^2$ for different values of SM4 parameters
both for the $B\rar K^\ast \mu^+ \mu^-$ and $B\rar K^\ast \tau^+
\tau^-$ decay.
\end{itemize}

Before performing further numerical analysis, few words about
lepton polarizations are in order. From explicit expressions of
the lepton polarizations one can easily see that they depend on
both $\ss$ and the new parameters($m_{t'},\,r_{sb}$). We should
eliminate the dependence of the lepton polarization on one of the
variables. We eliminate the variable $\hat{s}$ by performing
integration over $\ss$ in the allowed kinematical region. The
total branching ratio and the averaged polarizations $FB$ are
defined as \bea {\cal B}_r&=&\ds \int_{4
m_\ell^2/m_{B}^2}^{(1-\sqrt{\hat{r}_{K^\ast}})^2}
 \frac{d{\cal B}}{d\hat{s}} d\hat{s},
\nnb\\\lla A^{ij}_{FB} \rra &=& \frac{\ds \int_{4
m_\ell^2/m_{B}^2}^{(1-\sqrt{\hat{r}_{K^\ast}})^2} A^{ij}
\frac{d{\cal B}}{d\hat{s}} d\hat{s}} {{\cal{B}}_r}~. \eea

\begin{itemize}

\item $\lla{\cal
A}_{FB}^{LL}\rra$  strongly depends on the fourth quark
mass($m_{t'}$)
 and sensitives to the  product of quark mixing matrix elements($r_{sb}$) for both $\mu$
 and $\tau$ channels(see Figs. 19--24). Furthermore, for both channels, $\lla{\cal
A}_{FB}^{LL}\rra$ is a
 decreasing function of both $m_{t'}$ and $r_{sb}$(see Figs.
 19--24). The situation for $\phi_{sb}=90^\circ$ is much more
 interesting. It starts from SM3 value(around $0.2$) and gets the minimum
 value of $\approx-0.2$ for both $\mu$
 and $\tau$ channels. The measurement of the magnitude and sign of
 the $\lla{\cal
A}_{FB}^{LL}\rra$ can be used as a good tool for searching SM4
families and NP beyond the SM.
\end{itemize}

From these analyses we can conclude that the measurement of the
magnitude, the sign and the zero position of   ${\cal A}_{FB}^{LL}$
and the measurement of the magnitude and the sign of $\lla{\cal
A}_{FB}^{LL}\rra$ asymmetries are an indication of the existence of
new physics beyond the SM.

To sum up, we presented the  analysis of the double lepton
polarization of forward--backward asymmetries and their averaged
values in the exclusive $B \rar K^\ast \ell^- \ell^+$ decay, by
using the SM with four generations of quarks. The sensitivity of the
double lepton polarization of forward--backward asymmetries and
their averaged values on the new parameters that come out of fourth
generations are studied. We found out that both polarized $FB$
asymmetries and their averaged values  depict a strong dependence on
the fourth quark ($m_{t'}$) and the product of quark mixing matrix
elements ($V_{t^\prime b}^\ast V_{t^\prime
s}=r_{sb}e^{i\phi_{sb}}$). We obtained that the study of the
magnitude and the sign of the polarized $FB$ asymmetries for both
$\mu$ and $\tau$ cases and zero position of the polarized $FB$ for
$\mu$ case can serve as a good tool to look for physics beyond the
SM. More precisely, the results can be used to indirect research to
look for the fourth generation of quarks.

\section{Acknowledgment}
The authors would like to thank T. M. Aliev for his useful
discussions.

\newpage

\section*{Figure captions}

{\bf Fig. (1)} The dependence of ${\cal A}_{FB}^{LL}$ on $q^2$ for
the $B\rar K^\ast \mu^+ \mu^-$ at four fixed values of
$m_{t'}:300, 400, 500, 600$ GeV,
$\phi_{sb}=60^\circ$ and $r_{sb}=0.01$.\\ \\
{\bf Fig. (2)} The same as in Fig. (1), but for the $r_{sb}=0.02$.\\ \\
{\bf Fig. (3)} The same as in Fig. (1), but for the $r_{sb}=0.03$.\\ \\
{\bf Fig. (4)}
The same as in Fig. (1), but for the $\tau$ lepton.\\
\\
 {\bf Fig. (5)} The same as in Fig. (2), but for the $\tau$ lepton.\\
 \\
{\bf Fig. (6)} The same as in Fig. (3), but for the $\tau$ lepton.\\
 \\
 {\bf Fig. (7)}The same as in Fig. (1), but for the $\phi_{sb}=90^\circ$.\\
 \\
{\bf Fig. (8)} The same as in Fig. (7), but for the $r_{sb}=0.02$.\\ \\
{\bf Fig. (9)} The same as in Fig. (7), but for the $r_{sb}=0.03$.\\ \\
{\bf Fig. (10)} The same as in Fig. (7), but for the $\tau$ lepton.\\
\\
 {\bf Fig. (11)} The same as in Fig. (8), but for the $\tau$ lepton.\\
 \\
{\bf Fig. (12)} The same as in Fig. (9), but for the $\tau$ lepton.\\
 \\
 {\bf Fig. (13)}The same as in Fig. (1), but for the $\phi_{sb}=120^\circ$.\\
 \\
{\bf Fig. (14)} The same as in Fig. (13), but for the $r_{sb}=0.02$.\\ \\
{\bf Fig. (15)} The same as in Fig. (13), but for the $r_{sb}=0.03$.\\ \\
{\bf Fig. (16)} The same as in Fig. (13), but for the $\tau$ lepton.\\
\\
 {\bf Fig. (17)} The same as in Fig. (14), but for the $\tau$ lepton.\\
 \\
{\bf Fig. (18)} The same as in Fig. (15), but for the $\tau$ lepton.\\
 \\
 {\bf Fig. (19)}The dependence of $\lla{\cal
A}_{FB}^{LL}\rra$ on $m_{t'}$, for the $B\rar K^\ast \mu^+ \mu^-$
at three fixed values of
$\phi_{sb}:60^\circ, 90^\circ, 120^\circ$ and $r_{sb}=0.01$.\\ \\
{\bf Fig. (20)} The same as in Fig. (19), but for the $r_{sb}=0.02$.\\ \\
{\bf Fig. (21)} The same as in Fig. (19), but for the $r_{sb}=0.03$.\\ \\
{\bf Fig. (22)} The same as in Fig. (19), but for the $\tau$ lepton.\\
\\
 {\bf Fig. (23)} The same as in Fig. (20), but for the $\tau$ lepton.\\
 \\
{\bf Fig. (24)} The same as in Fig. (21), but for the $\tau$ lepton.\\
 \\
\newpage

\begin{figure}
\vskip 1.5 cm
    \includegraphics{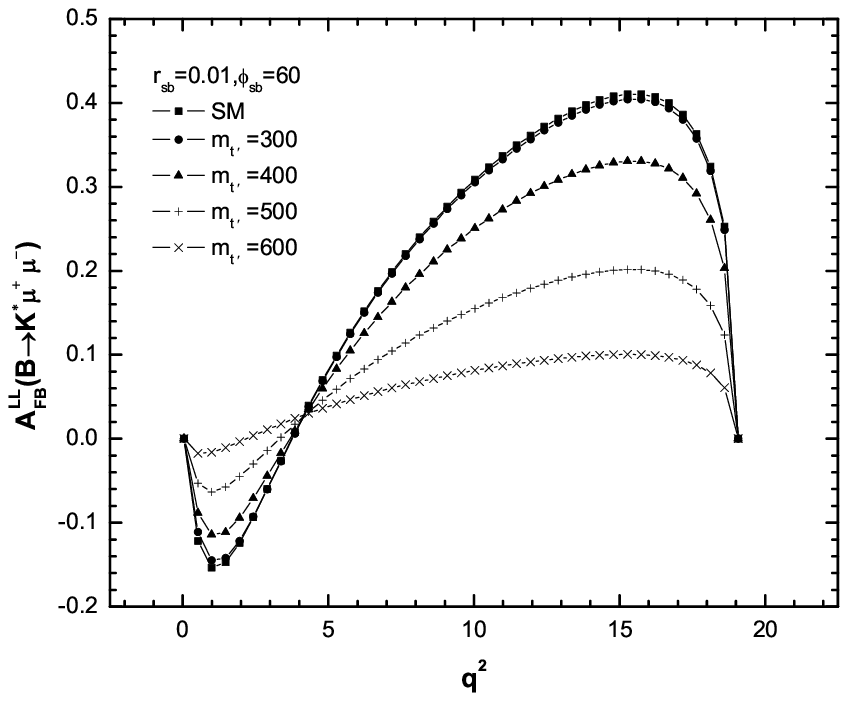}
\vskip 7.8cm \caption{}
\end{figure}

\begin{figure}
\vskip 2.5 cm
    \includegraphics{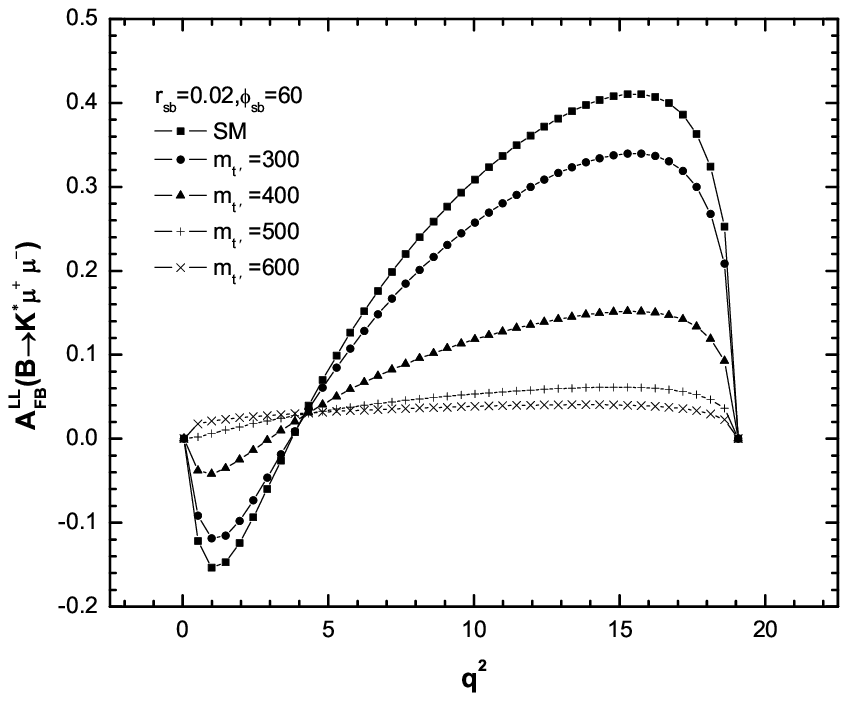}
\vskip 7.8 cm \caption{}
\end{figure}

\begin{figure}
\vskip 1.5 cm
    \includegraphics{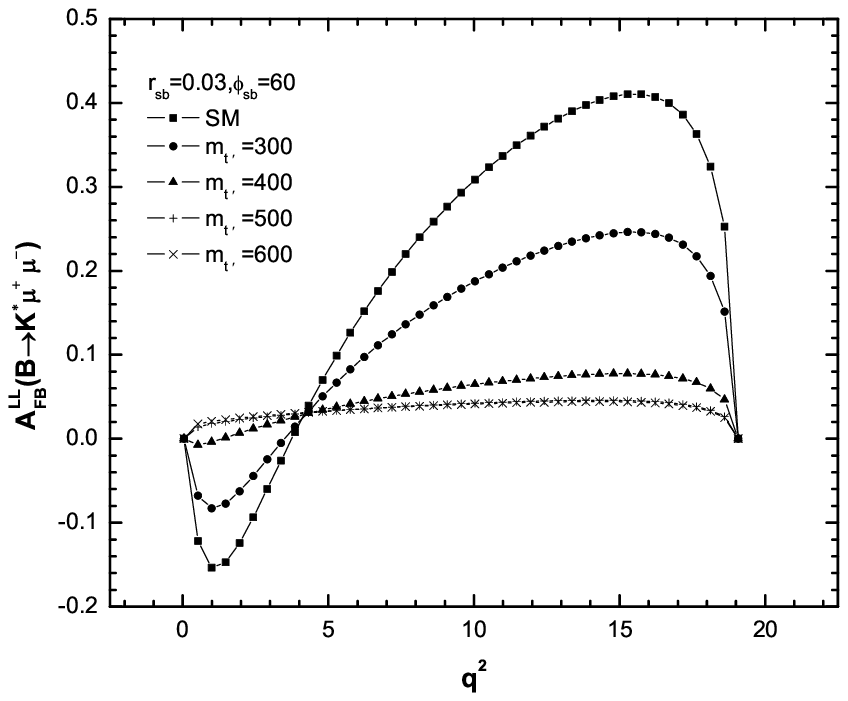}
\vskip 7.8cm \caption{}
\end{figure}

\begin{figure}
\vskip 2.5 cm
    \includegraphics{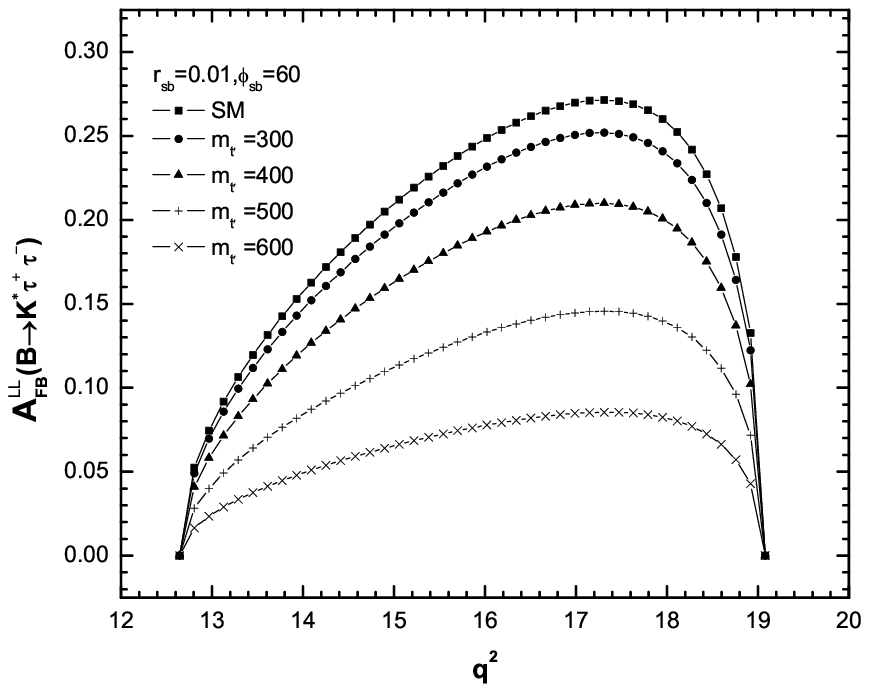}
\vskip 7.8 cm \caption{}
\end{figure}

\begin{figure}
\vskip 1.5 cm
    \includegraphics{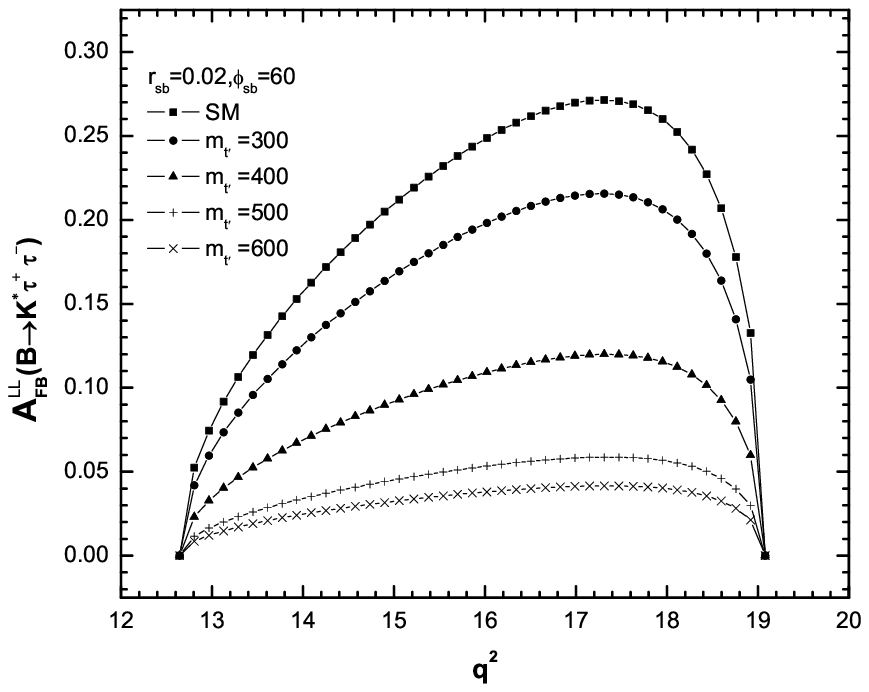}
\vskip 7.8cm \caption{}
\end{figure}

\begin{figure}
\vskip 2.5 cm
    \includegraphics{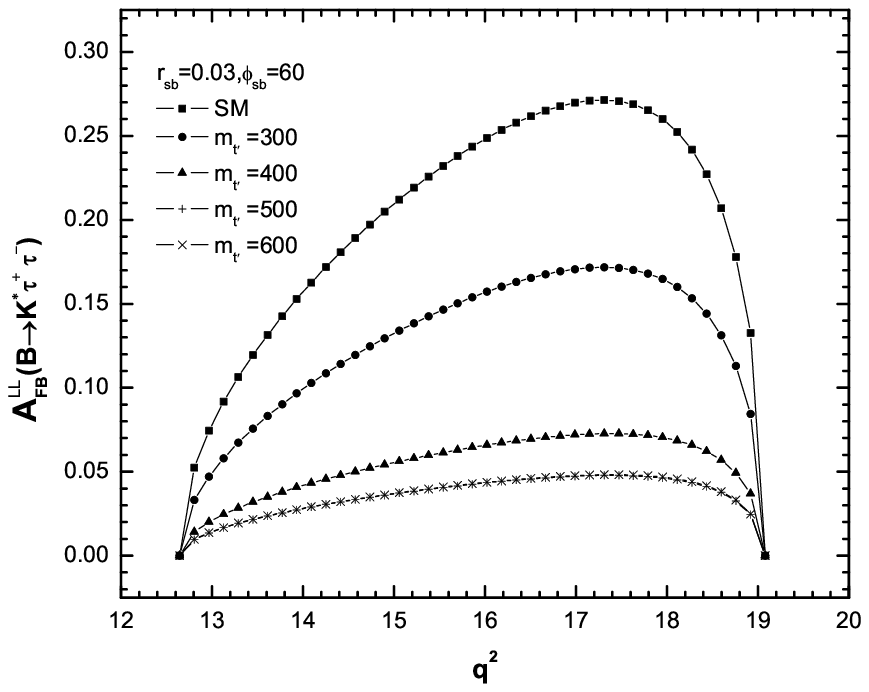}
\vskip 7.8 cm \caption{}
\end{figure}

\begin{figure}
\vskip 1.5 cm
    \includegraphics{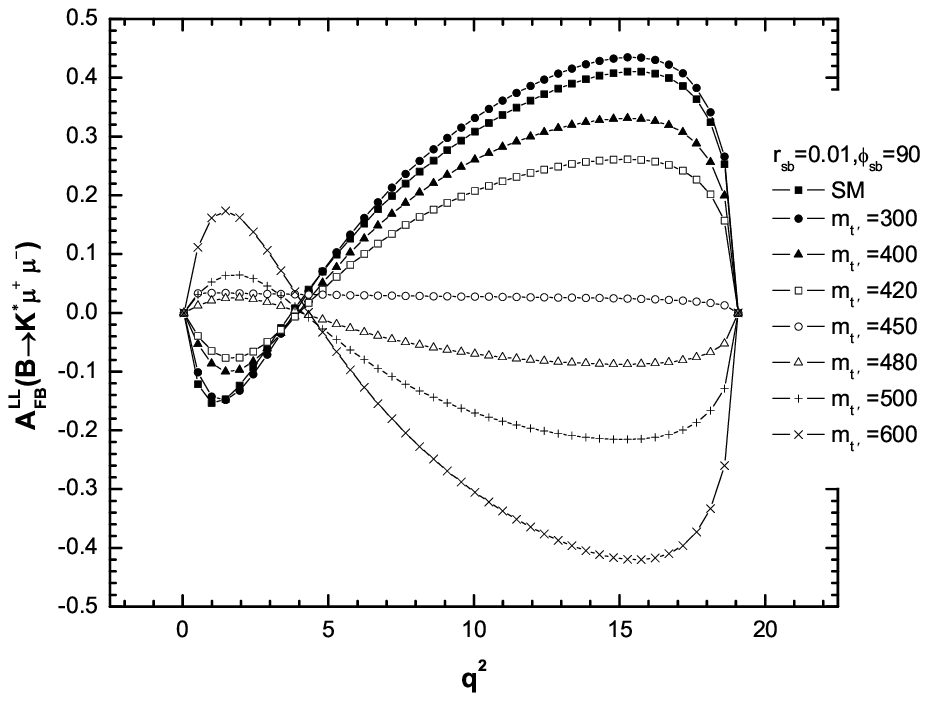}
\vskip 7.8cm \caption{}
\end{figure}

\begin{figure}
\vskip 2.5 cm
    \includegraphics{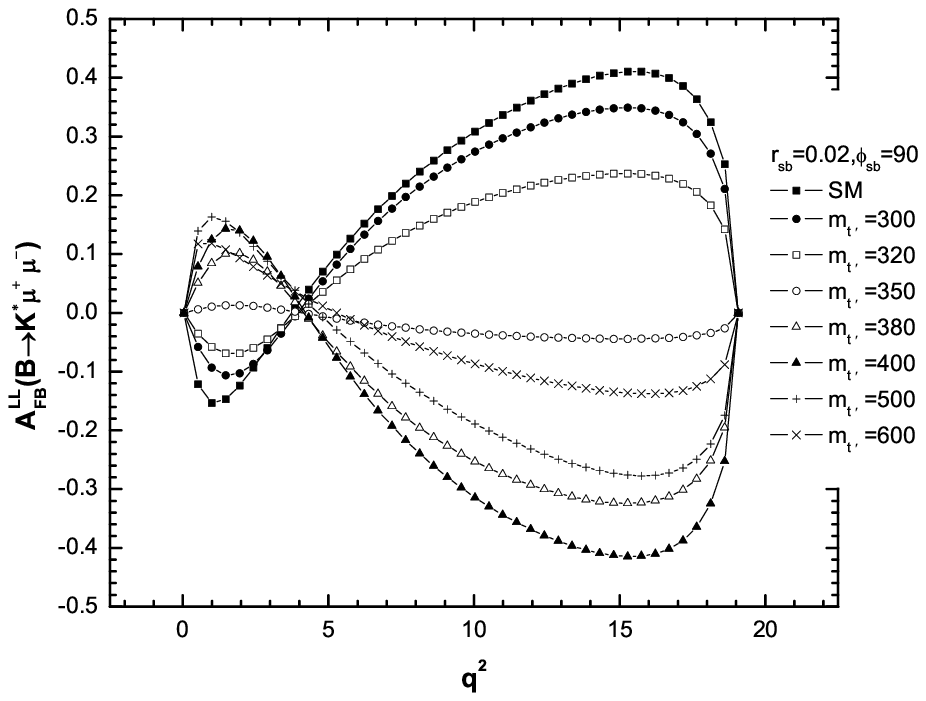}
\vskip 7.8 cm \caption{}
\end{figure}

\begin{figure}
\vskip 1.5 cm
    \includegraphics{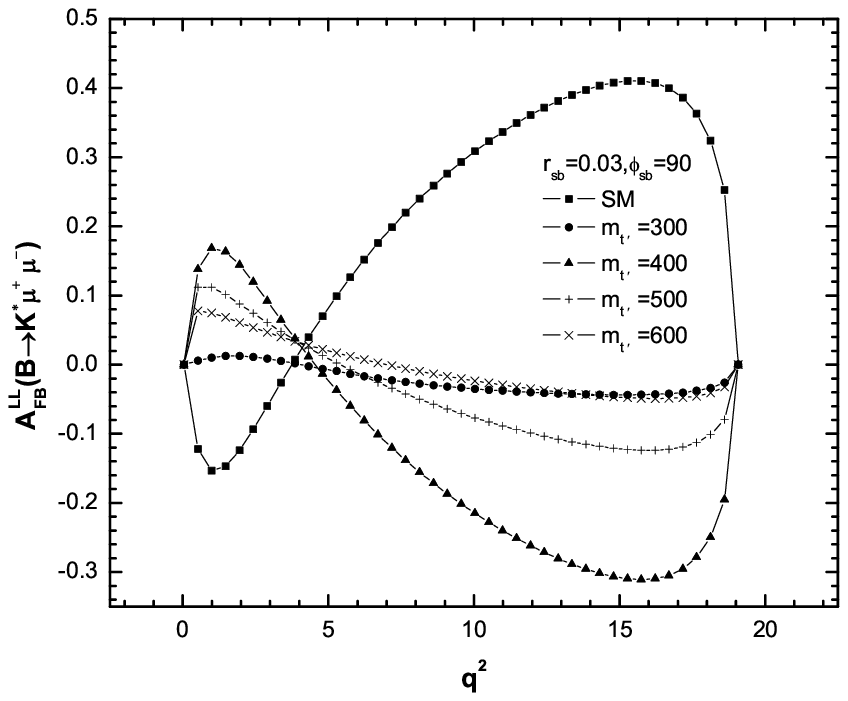}
\vskip 7.8cm \caption{}
\end{figure}

\begin{figure}
\vskip 2.5 cm
    \includegraphics{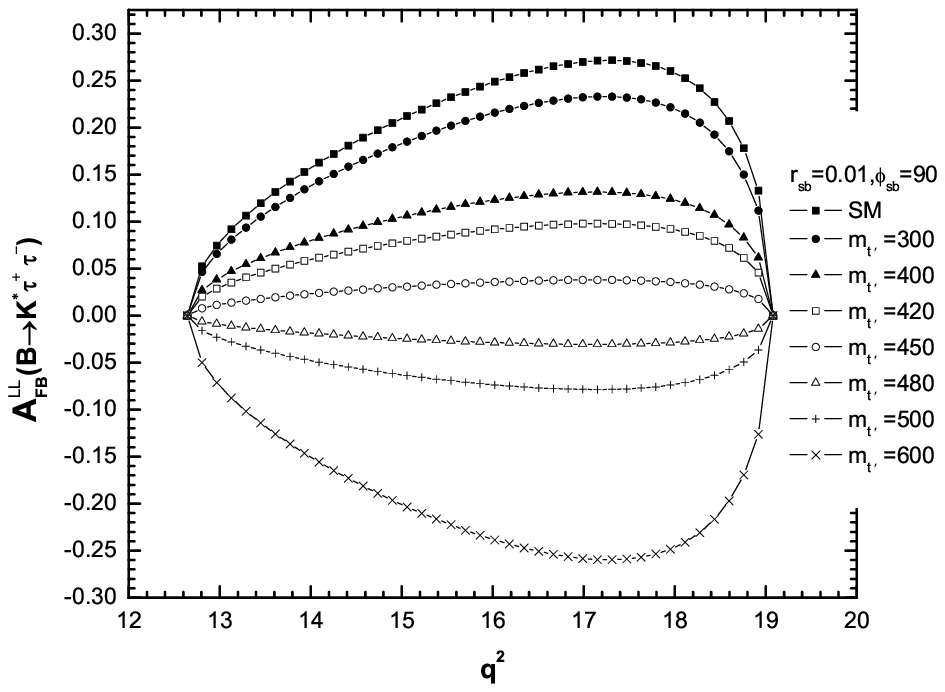}
\vskip 7.8 cm \caption{}
\end{figure}

\begin{figure}
\vskip 1.5 cm
    \includegraphics{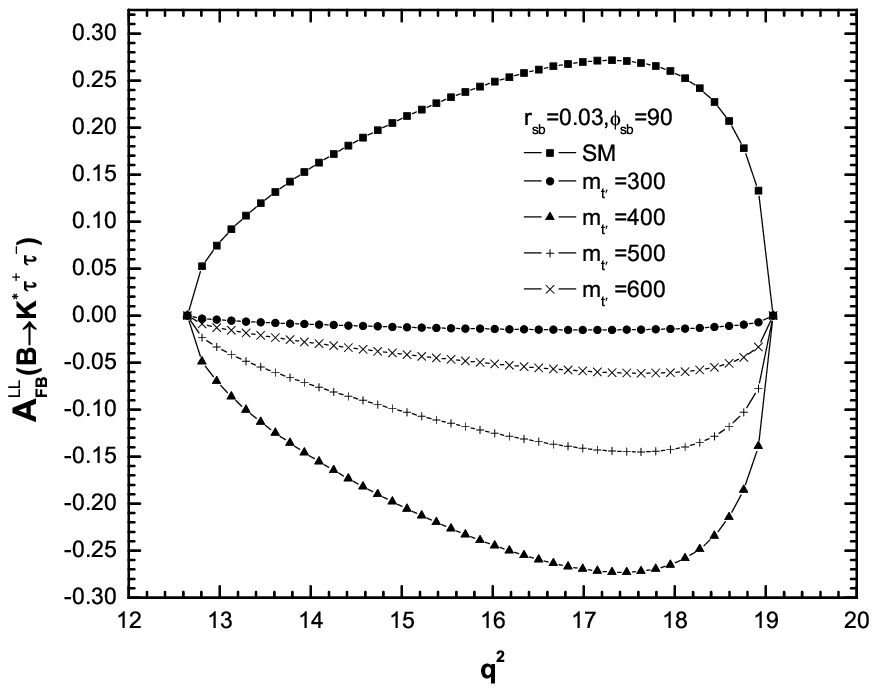}
\vskip 7.8cm \caption{}
\end{figure}

\begin{figure}
\vskip 2.5 cm
    \includegraphics{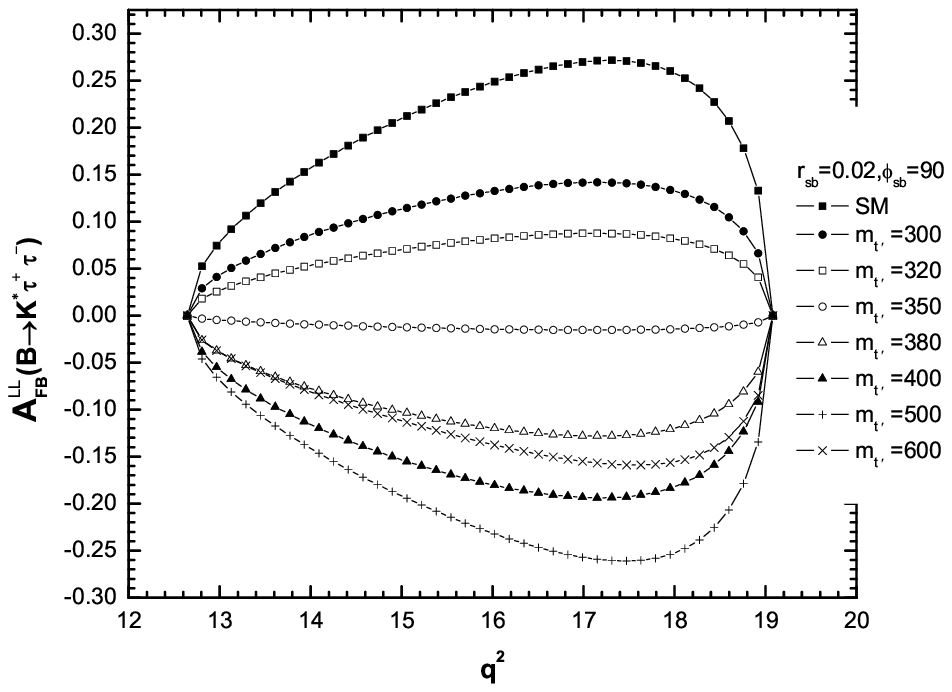}
\vskip 7.8 cm \caption{}
\end{figure}

\begin{figure}
\vskip 1.5 cm
    \includegraphics{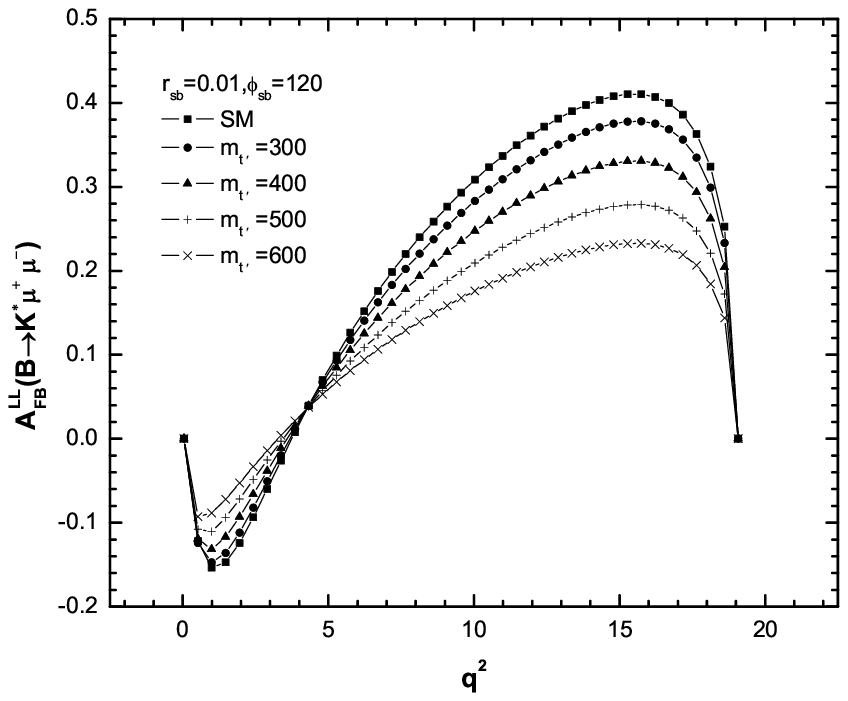}
\vskip 7.8cm \caption{}
\end{figure}

\begin{figure}
\vskip 2.5 cm
    \includegraphics{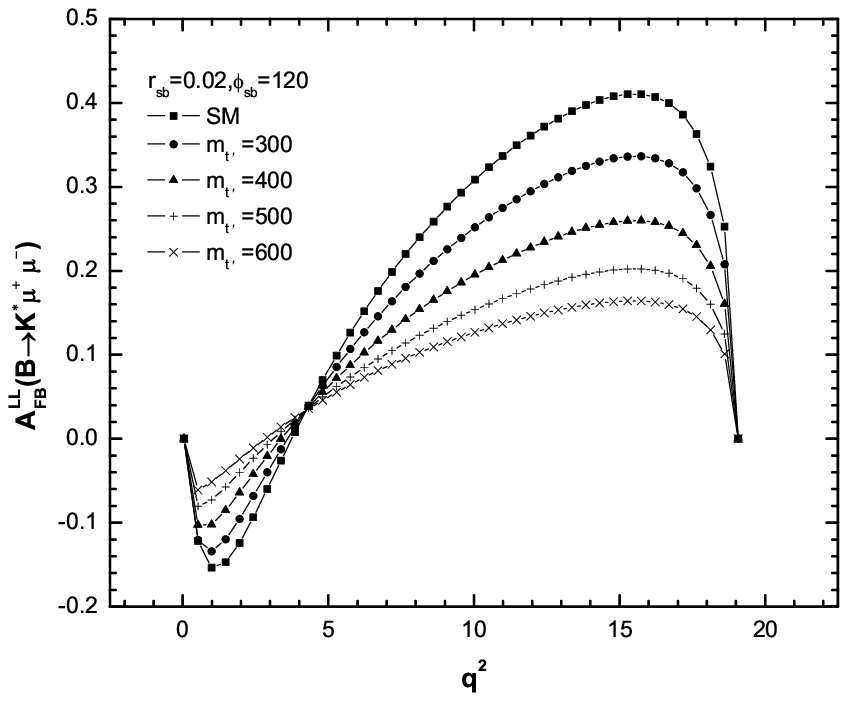}
\vskip 7.8 cm \caption{}
\end{figure}

\begin{figure}
\vskip 1.5 cm
    \includegraphics{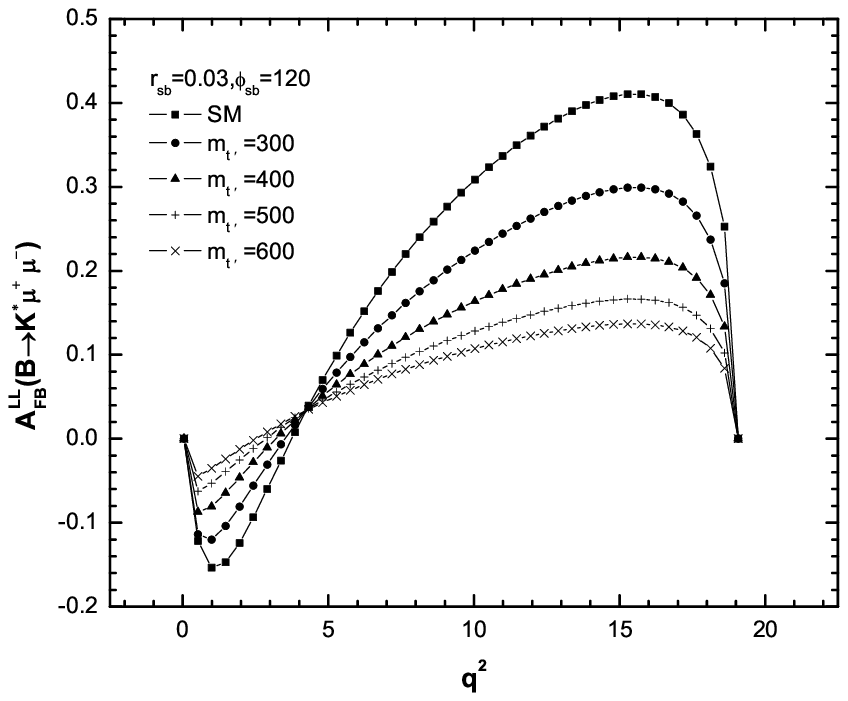}
\vskip 7.8cm \caption{}
\end{figure}

\begin{figure}
\vskip 2.5 cm
    \includegraphics{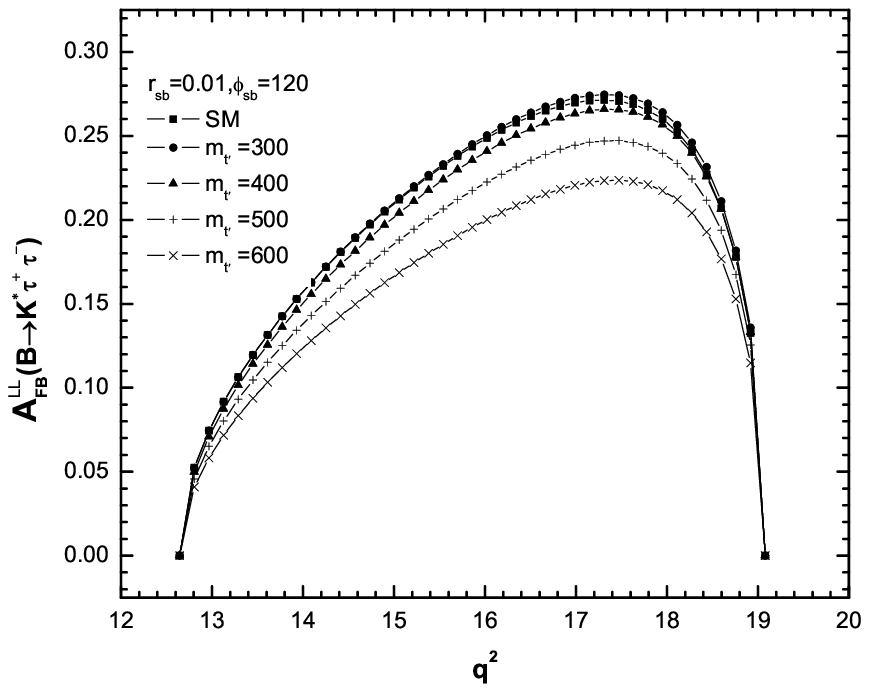}
\vskip 7.8 cm \caption{}
\end{figure}

\begin{figure}
\vskip 1.5 cm
    \includegraphics{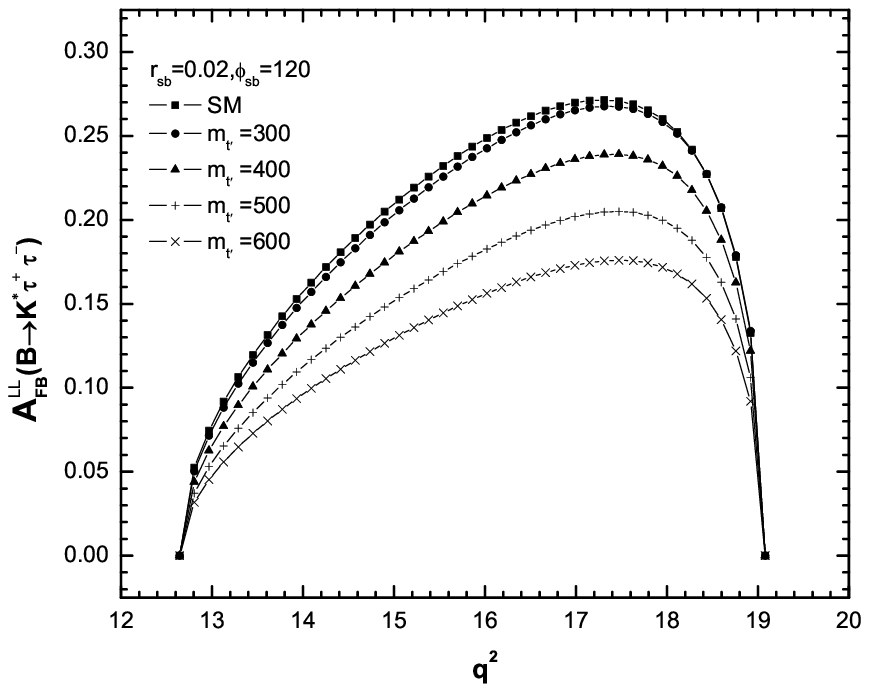}
\vskip 7.8cm \caption{}
\end{figure}

\begin{figure}
\vskip 2.5 cm
    \includegraphics{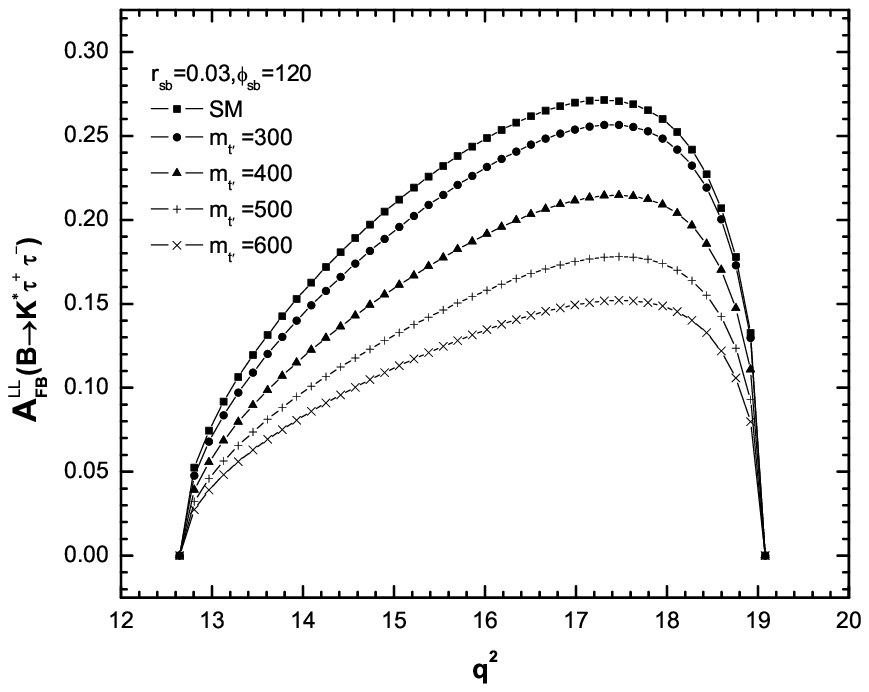}
\vskip 7.8 cm \caption{}
\end{figure}

\begin{figure}
\vskip 1.5 cm
    \includegraphics{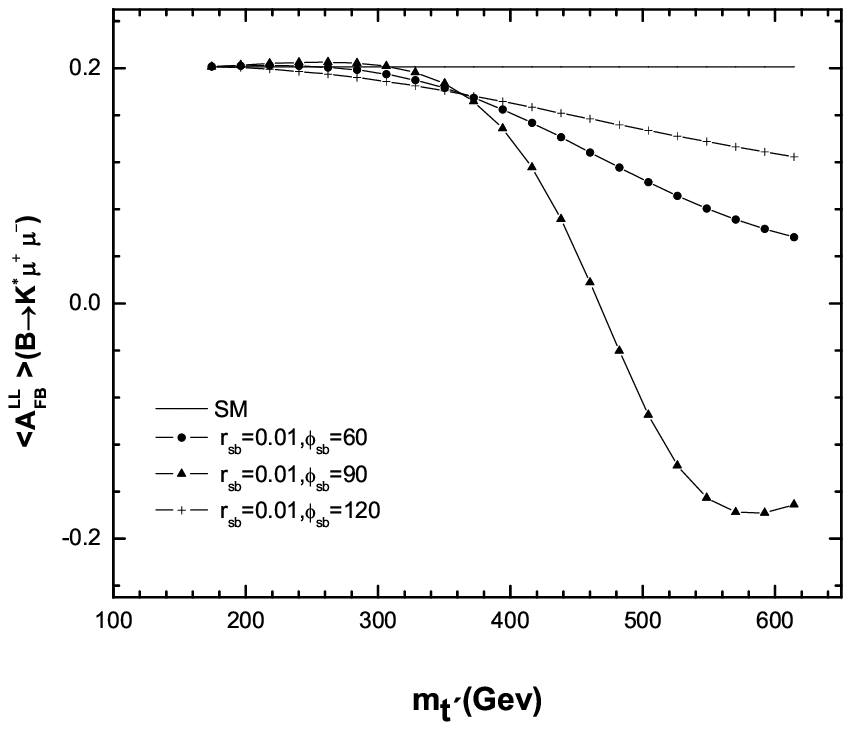}
\vskip 7.8cm \caption{}
\end{figure}

\begin{figure}
\vskip 2.5 cm
    \includegraphics{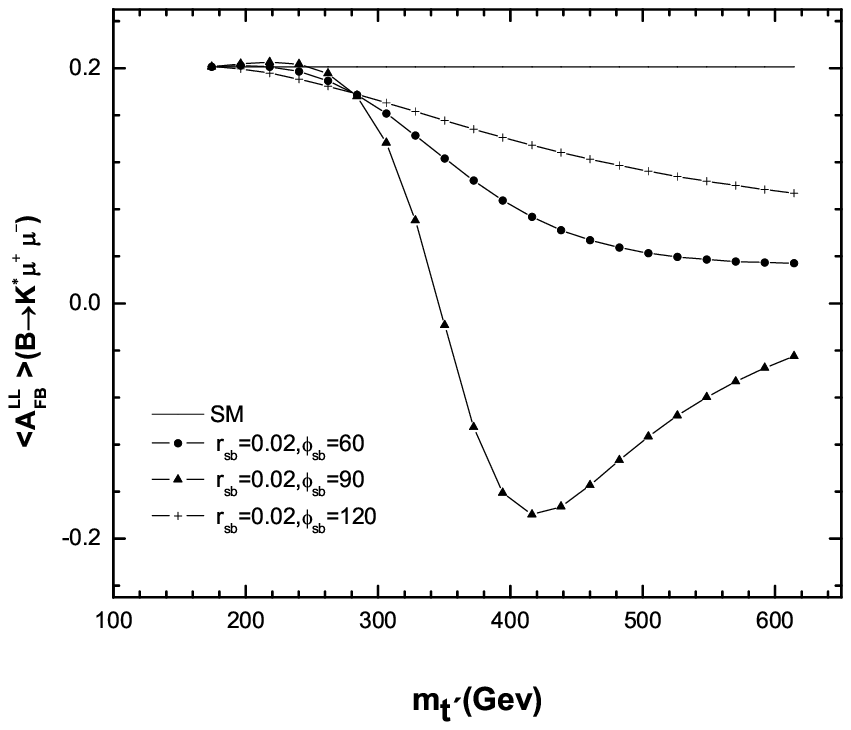}
\vskip 7.8 cm \caption{}
\end{figure}

\begin{figure}
\vskip 1.5 cm
    \includegraphics{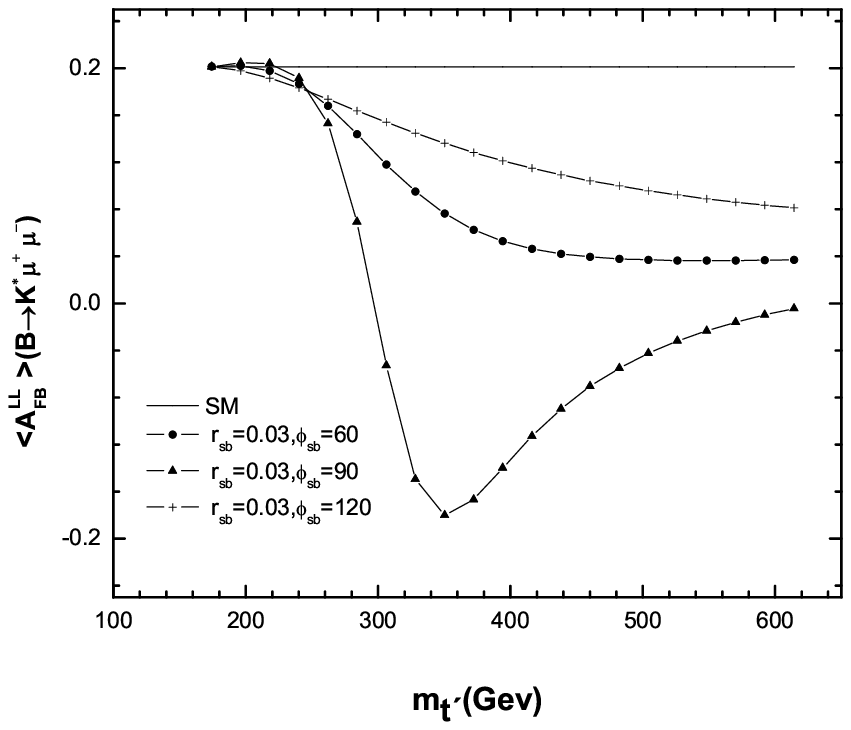}
\vskip 7.8cm \caption{}
\end{figure}

\begin{figure}
\vskip 2.5 cm
    \includegraphics{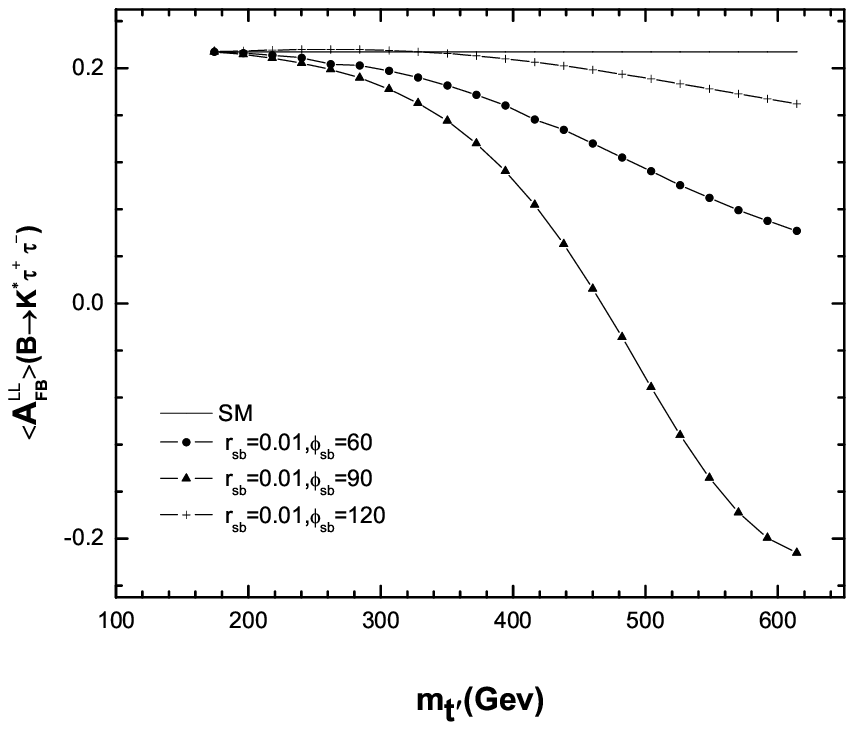}
\vskip 7.8 cm \caption{}
\end{figure}

\begin{figure}
\vskip 1.5 cm
    \includegraphics{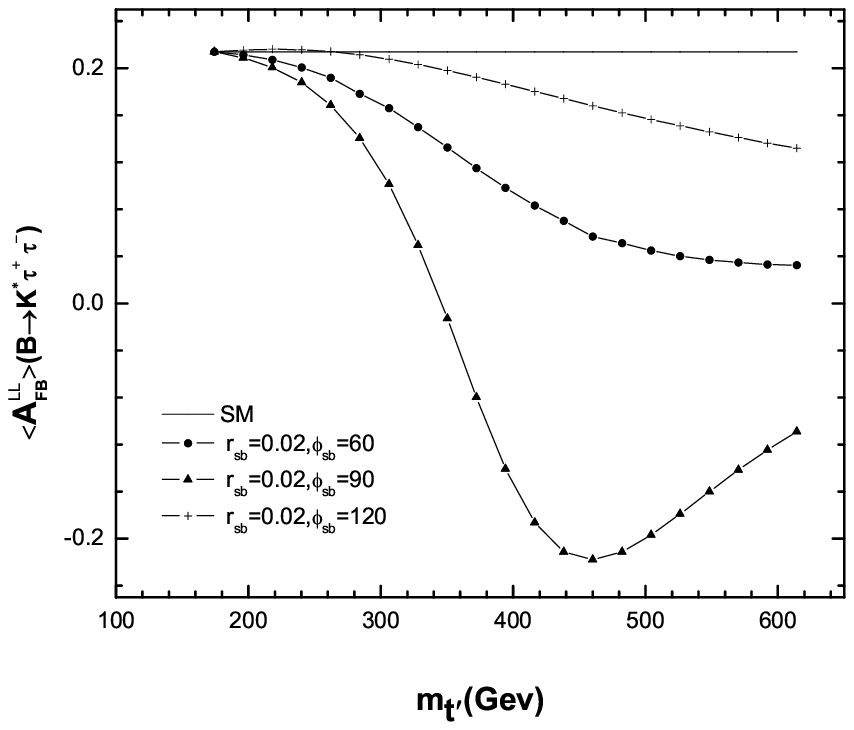}
\vskip 7.8cm \caption{}
\end{figure}

\begin{figure}
\vskip 2.5 cm
    \includegraphics{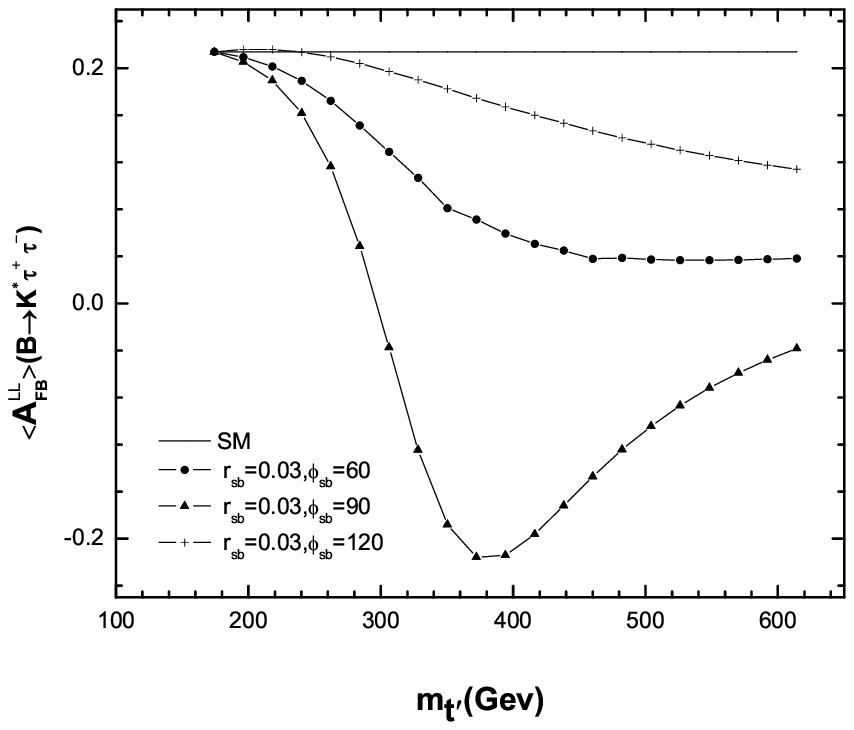}
\vskip 7.8 cm \caption{}
\end{figure}

\end{document}